\documentclass[aps,prb]{revtex4}
\usepackage{graphicx}
\usepackage{amsmath}

\numberwithin{equation}{section}

\begin{document}

\title{An introduction to the quasiclassical theory of superconductivity for diffusive proximity-coupled systems}
\author{Venkat Chandrasekhar}
\affiliation{Department of Physics and Astronomy, Northwestern University, Evanston, IL 60208, USA}
\date{\today}

\maketitle

Although the physics of normal metals (N) in close proximity to superconductors (S) has been studied extensively for more than thirty years, it is only in the past decade that experiments have been able to probe directly the region close to the NS interface at temperatures far below the transition temperature $T_c$ of the superconductor.  These experiments have been made possible by the availability of microlithography techniques that enable the fabrication of heterostructure devices with submicron scale features.  This size scale is comparable to the relevant physical length scales of the problem, and consequently, a number of new effects have been observed.  A variety of systems have been studied, the variation primarily being in the type of normal metal coupled to the superconductor.  Canonical normal metals such as copper and gold, semiconductors, insulators, and ferromagnets have been employed.  Although a variety of theoretical techniques have been used to describe proximity-coupled systems, the quasiclassical theory of superconductivity \cite{eilenberger,schmid,eliashberg,larkin,larkin2,hu,serene,schoen,aronov,beyer,rammer,smith} has proved to be a remarkably powerful tool in understanding the microscopic basis for the remarkable effects observed in these systems.  

A number of excellent recent articles \cite{volkov,lambert,volkov2,golubov,wilhelm,belzig,nazarov} have explored the application of the quasiclassical theory of superconductivity to proximity coupled systems.  In this review, a self-contained development of the quasiclassical theory is presented, starting from non-equilibrium Keldysh Green's functions for normal metal systems.  If the normal metal is clean, quasiparticles in the normal metal travel ballistically over long length scales; in the samples studied in the majority of recent experiments, however, the quasiparticles are scattered elastically within a short distance, so that the quasiparticle motion is diffusive.  Here, we shall concentrate on this diffusive case, where the elastic scattering length $\ell$ is the shortest relevant length scale in the problem.  We shall also restrict ourselves to the case where the superconductor is of the canonical $s$-wave type, avoiding complications with non-spatially symmetric order parameters within the superconductor.

\section{Transport equations in the diffusive approximation}
As an illustration of some of the issues that arise in dealing with non-equilibrium transport in mesoscopic diffusive systems, we consider first the classical Boltzmann equation in the diffusion approximation.  Consider a one-dimensional diffusive normal metal wire of length $L$.  We assume that all dimensions of the wire are larger than $\ell$, the elastic mean free path of the conduction electrons.  $f(E,x)$, the distribution function of electrons at energy $E$ and at a point $x$ along the wire, obeys the diffusion equation
\begin{equation}
D \frac{d^2 f(E,x)}{dx^2} + C(f)=0.
\label{eqn1.1}
\end{equation}
Here, $D=(1/3) v_F \ell $ is the three-dimensional electron diffusion constant, with $v_F$ being the Fermi velocity of the electrons. $C(f)$ is the collision integral, which takes into account inelastic scattering of the electrons, and itself depends on the distribution function $f$.  If we consider a mesoscopic wire whose length $L$ is much shorter than any inelastic scattering length $L_{in}$, this term can be set to  zero.

This diffusion equation has a simple solution in some specific geometries.  Consider the case of a wire of length $L$ sandwiched between two normal `reservoirs,' which we shall call the left (at $x=0$) and right (at $x=L$) reservoirs.  The reservoirs are defined as having an equilibrium distribution function $f_L(E)$ and $f_R(E)$ respectively.  Then the solution of the diffusion equation, Eqn.(\ref{eqn1.1}), under the condition $L \ll L_{in}$ (so that $C(f)=0$), and subject to the boundary conditions that $f(E,x)=f_L(E)$ at the left reservoir, and $f(E,x)=f_R(E)$ at the right reservoir, is given by \cite{nagaev}
\begin{equation}
f(E,x)=\left[f_R(E)-f_L(E)\right]\frac{x}{L} +f_L(E)
\label{eqn1.2}
\end{equation}
The electrical current $I$ through the wire in the diffusion approximation is given by
\begin{equation}
I=-e A D \int N(E) \frac{\partial f}{\partial x} dE
\label{eqn1.3}
\end{equation}
where $A$ is the cross-sectional area of the wire.  With $f(E,x)$ given by Eqn.(\ref{eqn1.2}), this can be written
 \begin{equation}
I=-\frac{eAD}{L}  \int N(E) [f_R(E)-f_L(E)] dE
\label{eqn1.4}
\end{equation}
In order to obtain a finite current, we must apply a voltage across the wire.  Let us apply a voltage $V$ to the left reservoir, keeping the right reservoir at ground ($V=0$).  This has the effect of shifting the energies of the electrons in the left reservoir by $-eV$, so that the electron distribution function there is given by $f_L(E)=f_0(E-eV)$, where $f_0(E)$ is the usual equilibrium Fermi distribution function
\begin{equation}
f_0(E)=\frac{1}{e^{E/k_B T} + 1}
\label{eqn1.5}
\end{equation}
and the energy $E$ is measured from the Fermi energy $E_F$.   Figure 1 shows the distribution function $f(E,x)$ in the wire.  It has a step-function form, which varies linearly along the length of the wire.  Evidence for such a non-equilibrium distribution has recently been seen experimentally in a series of beautiful experiments by the Saclay group.\cite{pierre}

\begin{figure}
\center{\includegraphics[width=10cm]{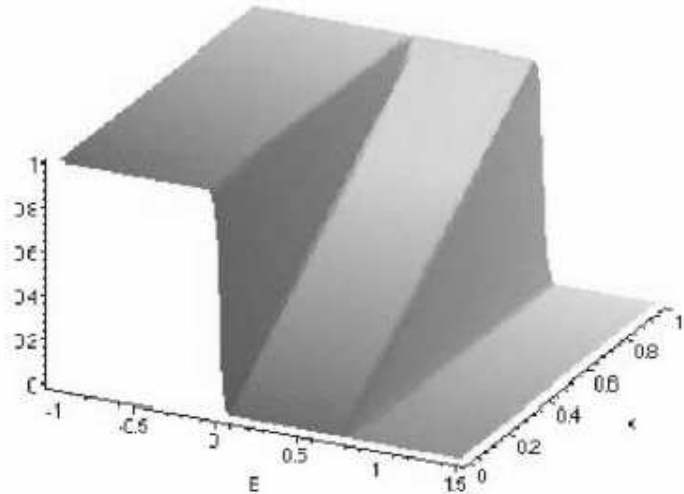}}
\caption{Nonequilibrium electron distribution function $f$ in a normal wire with a potential $V$ applied across it, as a function of energy $E$ and position $x$.  The position $x$ is normalized to the length $L$ of the wire.}
\label{fig1}
\end{figure}

In the limit of small voltages $V$, the difference of distribution functions in Eqn.(\ref{eqn1.4}) can be expanded as
\begin{equation}
f_R(E)-f_L(E)=f_0(E)-f_0(E-eV) \approx f_0(E)- \left[ f_0(E) + eV \left(-\frac{\partial f_0}{\partial E}\right)\right] = -eV\left(-\frac{\partial f_0}{\partial E}\right)
\label{eqn1.6}
\end{equation}
so that
\begin{equation}
I=\frac{e^2 A D}{L} V \int N(E)  \left(-\frac{\partial f_0} { \partial E} \right) dE \approx \frac{N(0) e^2 A D}{L} V.
\label{eqn1.7}
\end{equation}
where $N(0)$ is the density of states at the Fermi energy.  In performing the integral, we have assumed that we are at low enough temperatures so that the derivative of the Fermi function $(-\partial f_0 / \partial E)$ can be approximated by $\delta (E)$.  The electrical conductance is then given in the Nernst-Einstein form $G=N(0)e^2D (A/L)=\sigma_0 (A/L)$.

The thermal current through the wire is given by an expression similar to Eqn.({\ref{eqn1.4}):
\begin{equation}
I^T=\frac{AD}{L}  \int  E N(E) [f_R(E)-f_L(E)] dE
\label{eqn1.8}
\end{equation}
The critical difference between the expression for the thermal current and the expression for the electrical current is the presence of an additional factor of $E$ in the integrand in Eqn.(\ref{eqn1.8}), which has important consequences in the calculation of the thermal properties.

To obtain expressions for the thermal coefficents, let the temperature of the left reservoir be $T$ and its voltage $V$, and let the temperature of the right reservoir $T+\Delta T$ and its corresponding voltage $V=0$.  The distribution function in the left electrode is then $f_L(E)=f_0(E-eV, T)$, and the distribution function in the right electrode is $f_R(E)=f_0(E, T+\Delta T)$.    In the limit of small $\Delta T, V$, the difference in the distribution functions in Eqn.(\ref{eqn1.4}) can be expanded as
\begin{equation}
f_R(E)-f_L(E)= \left[ f_0(E,T) + \frac{E}{T}\Delta T \left(-\frac{\partial f_0}{\partial E} \right) \right] -  \left[ f_0(E,T) + eV \left(-\frac{\partial f_0}{\partial E} \right) \right] \\
= \left[ \frac{E}{T}\Delta T -eV \right]\left(-\frac{\partial f_0}{\partial E} \right)
\label{eqn1.9}
\end{equation}
Putting this expression into the equations for the electrical and thermal currents, Eqn.(\ref{eqn1.4}) and Eqn.(\ref{eqn1.8}), one obtains the two transport equations
\begin{subequations} \label{eqn1.10}
\begin{align}
I & =\frac{e^2 A D}{L} \left[ \int N(E)  \left(-\frac{\partial f_0} { \partial E} \right) dE \right] V  -  \frac{eAD}{TL}\left[ \int E N(E)  
\left(-\frac{\partial f_0} { \partial E} \right) dE \right] \Delta T  \label{eqn1.10a} \\
I^T &=-\frac{e A D}{L} \left[ \int E N(E)  \left(-\frac{\partial f_0} { \partial E} \right) dE \right] V  +  \frac{AD}{TL}\left[ \int E^2 N(E)  
\left(-\frac{\partial f_0} { \partial E} \right) dE \right] \Delta T  \label{eqn1.10b}.
\end{align}
\end{subequations}
These equations are equivalent to the usual form for the transport equations for a metal:
\begin{subequations} \label{eqn1.11}
\begin{align}
I &= GV + \eta \Delta T \label{eqn1.11a}  \\
I^T&=\zeta V + \kappa \Delta T \label{eqn1.11b}
\end{align}
\end{subequations}
with the linear-response thermoelectric coefficients defined by 
\begin{subequations} \label{eqn1.12}
\begin{align}
G &=\frac{e^2 A D}{L} \left[ \int N(E)  \left(-\frac{\partial f_0} { \partial E} \right) dE \right], \label{eqn1.12a}  \\
\eta &=-  \frac{eAD}{TL}\left[ \int E N(E)  
\left(-\frac{\partial f_0} { \partial E} \right) dE \right] ,
\label{eqn1.12b} \\
\zeta &=-\frac{e A D}{L} \left[\int E N(E)  \left(-\frac{\partial f_0} { \partial E} \right) dE \right], 
\label{eqn1.12c} \\
\kappa &=\frac{AD}{TL}\left[ \int E^2 N(E)  \left(-\frac{\partial f_0} { \partial E} \right) dE \right].
\label{eqn1.12d}
\end{align}
\end{subequations}
If we approximate the derivative of the Fermi function by a $\delta$ function at low temperatures, as we did for the electrical conductance $G$, we see that all other coefficients vanish.  In order to obtain a finite value, we use the Sommerfeld expansion for the integrals \cite{ziman}
\begin{equation}
\int \Phi(E) \left(-\frac{\partial f_0}{\partial E} \right) dE = \Phi(0) + 
	\frac{\pi^2 (k_B T)^2}{6} \left[\frac{\partial^2 \Phi(E)}{\partial E^2}\right]_{E=0} + \ldots
\label{eqn1.13}
\end{equation}
Using the fact that $N'(0)=N(0)/2 E_F$, we obtain the following expressions for the last three coefficients
\begin{subequations} \label{eqn1.14}
\begin{align}
\eta&=-\frac{\pi^2 k_B^2 T}{6} \frac{eAD}{L}\frac{N(0)}{E_F},
\label{eqn1.14a} \\
\zeta&=-\frac{\pi^2 (k_B T)^2}{6} \frac{eAD}{L}\frac{N(0)}{E_F},
\label{eqn1.14b} \\
\kappa&=\frac{\pi^2 k_B^2 T}{3} \frac{AD}{L}N(0).
\label{eqn1.14c}
\end{align}
\end{subequations}
Experimentally, the quantities often measured are the thermopower $S$ and the thermal conductance $G^T$.  The thermopower is defined as the thermal voltage generated by a temperature differential $\Delta T$, under the condition that no electrical current flows through the wire ($I=0$).  Putting this condition into the first transport equation, Eqn.({\ref{eqn1.10a}), we obtain
\begin{equation}
S=\frac{V}{\Delta T}=-\frac{\eta}{G}=\frac{\pi^2}{6}\frac{k_B}{e}\frac{k_B T}{E_F}
\label{eqn1.15}
\end{equation}
Note that since the expression for $S$ contains a factor $k_B T/E_F$, the thermopower of a typical normal metal is very small.

The thermal conductance $G^T$ is defined as the ratio of the thermal current $I^T$ to the temperature differential $\Delta T$, under the same condition of no electrical current flow ($I=0$).  From the equations above
\begin{equation}
G^T=\frac{I^T}{\Delta T}=\kappa + \zeta S
\label{eqn1.16}
\end{equation}
For typical metals, the second term is much smaller than the first, and is usually ignored, so that $G^T \approx \kappa$.  If we take the ratio of the electrical to the thermal conductance, we obtain
\begin{equation}
\frac{G^T}{G}= \frac{\pi^2}{3}\frac{k_B^2}{e^2} T. 
\label{eqn1.17}
\end{equation}
Consequently, one finds that the Wiedemann-Franz Law holds, even though the scattering lengths for momentum relaxation ($\ell$) and energy relaxation ($L_{in}$) are quite different.  This is because the reservoirs are assumed to be perfect, so that any electron entering a reservoir immediately equilibrates with the other electrons in the reservoir.

Before we go on to discuss normal metals in contact with superconductors, it is worthwhile to review some important assumptions in the calculations above.  First, in our calculations, we have assumed that the diffusion coefficient $D$ is a constant independent of the energy $E$ and position $x$.  When coherence effects are important, as in the case of the proximity effect, the diffusion coefficient becomes a function of both these parameters, $D(E,x)$.  The differential equation for the distribution function, Eqn.(\ref{eqn1.1}) is then modified.  $D(E,x)$  itself in general is determined by the distribution function $f(E,x)$, so one must solve a set of coupled differential equations to obtain a solution.  This is difficult to do analytically in all but the simplest of cases, and more often must be done numerically.  The remainder of this chapter will be devoted in great part to deriving the appropriate differential equations for the distribution functions and diffusion coefficient in the case of normal metals in contact with superconductors, using the quasiclassical equations for superconductivity.

Second, apart from the electrical conductance $G$, the thermal coefficients derived above would all vanish if the density of states at the Fermi energy $N(E)$, were assumed to be constant, so that it could be taken out of the integrals.  The small variation in the density of states at the Fermi energy is responsible for finite  (but small) values of the off-diagonal transport coefficients $\eta$ and $\zeta$.  For example, the thermopower $S$, which is non-zero only if there is an asymmetry between the properties of particles and holes near the Fermi energy, vanishes if $N(E)$ is assumed constant.  The small difference in $N(E)$ above and below $E_F$ gives rise to the small but finite thermopowers of typical normal metals.  The conventional quasiclassical theory of superconductivity assumes particle-hole symmetry \textit{a priori} in the definition of the quasiclassical Green's functions, in that $N(E)$ is assumed constant at $E_F$.  Consequently, thermoelectric effects cannot be calculated in the conventional quasiclassical approximation; an extension of the theory is required.
 
Finally, we have been discussing here \textit{currents} and \textit{conductances}, rather than \textit{current densities} and \textit{conductivities}.  These are the more relevant quantitites, since we will be discussing mesoscopic samples in which the measured properties are properties of the sample as a whole.  This will be particularly important for the proximity effect, where long range phase coherence means that non-local effects are important.

\section{The Keldysh Green's Functions}
The starting point for developing the quantum analog to the classical Boltzmann transport equation is the Keldysh diagrammatic technique.  We shall begin our discussion of the Keldysh technique in the notation of Lifshitz and Pitaevskii.\cite{lifshitz}  As in the equilibrium case, we define a non-equilibrium Green's function 
\begin{equation}
G_{\sigma_1 \sigma_2}(X_1,X_2) =- i \left< n\left|T \hat{\psi}_{\sigma_1}(X_1) \hat{\psi}_{\sigma_2}^+(X_2)\right|n\right>
\label{eqn2.1}
\end{equation}
Here $X_1$ and $X_2$ take into account the three spatial coordinates (denoted by $\vec{r}_1$ and $\vec{r}_2$ respectively), and the time coordinate ($t_1$ and $t_2$).  The difference between the non-equilibrium Keldysh Green's function above and the usual equilibrium Green's function is that the average is taken over any quantum state $|n>$ of the system, rather than just the ground state $|0>$.  The time ordering operator $T$ has the effect
\begin{equation}
G_{\sigma_1 \sigma_2}(X_1,X_2) = 
\begin{cases}
 -i \left< n\left|\hat{\psi}_{\sigma_1}(X_1) \hat{\psi}_{\sigma_2}^+(X_2)\right|n\right> & \text{if $t_1>t_2$}, \\
+i \left< n\left| \hat{\psi}_{\sigma_2}^+(X_2)\hat{\psi}_{\sigma_1}(X_1)\right|n\right> & \text{if $t_2>t_1$}
\end{cases} 
\label{eqn2.2}
\end{equation}
for fermion operators (the case of interest here).

For simplicity, we label the spin indices and the coordinate indices by numbers.  We shall define a set of Green's functions that will be useful later:
 \begin{subequations} \label{eqn2.3}
 \begin{align}
G_{12}^{\alpha \alpha} &= -i \left<T \hat{\psi}_1 \hat{\psi}_2^+ \right>,
\label{eqn2.3a} \\
G_{12}^{\beta \beta} &= i \left<\tilde{T} \hat{\psi}_1^+ \hat{\psi}_2 \right>,
\label{eqn2.3b} \\
G_{12}^{\alpha \beta} &= -i \left<\hat{\psi}_1 \hat{\psi}_2^+ \right>,
\label{eqn2.3c} \\
G_{12}^{\beta \alpha} &= i \left< \hat{\psi}_2^+ \hat{\psi}_1 \right>.
\label{eqn2.3d}
\end{align}
\end{subequations}
The first Green's function is just the one defined above, in this new notation.  The second Green's function is similar in definition to $G_{12}^{\alpha \alpha}$, except that it is defined with operator $\tilde{T}$ instead of $T$, which orders the operators in reverse chronological order.  The last two Green's functions are defined without any time-ordering operators.  The four Green's functions so defined are not linearly independent,  but are related by linear equations of the form
\begin{equation}
G^{\alpha \alpha} + G^{\beta \beta} = G^{\alpha \beta} + G^{\beta \alpha}.
\label{eqn2.4}
\end{equation}
The retarded and advanced Green's functions $G^R$ and $G^A$ can be defined as in the equilibrium case
\begin{equation}
G_{12}^R=
\begin{cases}
-i \left<\hat{\psi}_1 \hat{\psi}_2^+ +  \hat{\psi}_2^+\hat{\psi}_1 \right> & \text{if $t_1>t_2$}, \\
0 & \text{if $t_2>t_1$}
\end{cases}
\label{eqn2.5}
\end{equation}
and
\begin{equation}
G_{12}^A=
\begin{cases}
0 & \text{if $t_1>t_2$},  \\
i \left<\hat{\psi}_1 \hat{\psi}_2^+ +  \hat{\psi}_2^+\hat{\psi}_1 \right> & \text{if $t_2>t_1$}.
\end{cases}
\label{eqn2.6}
\end{equation}
and are related by 
\begin{equation}
G_{12}^A=\left(G_{21}^R\right)^*.
\label{eqn2.7}
\end{equation}
$G^R$ and $G^A$ can be written in terms of the Keldysh Green's functions defined earlier as
\begin{subequations} \label{eqn2.8}
\begin{align}
G^R &=G^{\alpha \alpha} - G^{\alpha \beta} = G^{\beta \alpha} - G^{\beta \beta} \label{eqn2.8a} \\
G^A &=G^{\alpha \alpha} - G^{\beta \alpha } = G^{ \alpha \beta} - G^{\beta \beta} \label{eqn2.8b}
\end{align}
\end{subequations}
$G^{\alpha \alpha}$ satisfies the equation of motion
\begin{equation}
G_{01}^{-1} G_{12}^{(0)\alpha \alpha}= \delta(X_1 - X_2)
\label{eqn2.9}
\end{equation}
where $G_0^{-1}$ is the differential operator (in the free electron approximation)\footnote{We set $\hbar=1$.}
\begin{equation}
G_0^{-1} = i\frac{\partial}{\partial t} + \frac{\nabla^2}{2 m} + \mu
\label{eqn2.10}
\end{equation}
and the second subscript to $G_0^{-1}$ denotes that the differentials in Eqn.(\ref{eqn2.10}) are with respect to coordinates corresponding to this subscript.  The argument of the delta function Eqn.(\ref{eqn2.9}) includes space, time and spin coordinates, and the notation $G_{12}^{(0)\alpha \alpha}$ (with a superscript 0) signifies a Green's function for an ideal gas.

The $\delta$ function in Eqn.(\ref{eqn2.9}) arises from the discontinuities in $G^{\alpha \alpha}$ at $t_1=t_2$.  $G^R$ and $G^A$ have similar discontinuities, and obey a similar equation.  $G^{(0)\beta \beta}$ has a discontinuity of the opposite sign, and hence obeys the equation
\begin{equation}
G_{01}^{-1} G_{12}^{(0)\beta \beta}= -\delta(X_1 - X_2)
\label{eqn2.11}
\end{equation}
$G^{\alpha \beta}$ and $G^{\beta \alpha }$  have no discontinuities at $t_1=t_2$, and hence obey the equations
\begin{subequations} \label{eqn2.12}
\begin{align}
G_{01}^{-1} G_{12}^{(0)\beta \alpha} &= 0 \label{eqn2.12a} \\
G_{01}^{-1} G_{12}^{(0) \alpha \beta} &= 0 \label{eqn2.12b} 
\end{align}
\end{subequations}
   
The diagram technique for Keldysh Green's functions is similar to that for equilibrium Green's  functions, except that one needs to sum over the internal indices $\alpha$ and $\beta$, with a corresponding increase in the number of diagrams.  For example, Fig.1 shows two diagrams corresponding to the first order corrections to $G_{12}^{\alpha \alpha}$ in the presence of an external potential, which is represented by a dashed line. 

\begin{figure}
\center{\includegraphics[width=8cm]{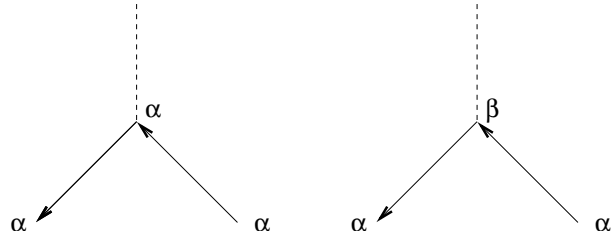}}
\caption{Diagrams corresponding to the first order corrections to the Keldysh Green's function $G_{12}^{\alpha \alpha}$ in the presence of an external potential $U$, which is represented by a dashed line.}
\label{fig2}
\end{figure}
In the same manner, Dyson's equation for the Green's function can be represented as shown in Fig. 2, where the ellipse represents the self-energy function $\Sigma^{\alpha \beta}$.  In order to make the notation more compact and tractable, it is useful to introduce a matrix Green's function and corresponding self-energy matrix:
\begin{equation}
\check{G}=
\begin{pmatrix}
G^{\beta \beta} &   G^{\alpha \beta}  \\
G^{\beta \alpha} &   G^{\alpha \alpha}
\end{pmatrix}, 
\qquad
\check{\Sigma}=
\begin{pmatrix}
\Sigma^{\beta \beta} &   \Sigma^{\alpha \beta}  \\
\Sigma^{\beta \alpha} &   \Sigma^{\alpha \alpha}
\end{pmatrix}.
\label{eqn2.13}
\end{equation}
Dyson's equation can then be written in matrix form as
\begin{equation}
\check{G}_{12} = \check{G}_{12}^0 + \int \check{G}_{14}^0 \check{\Sigma}_{43} \check{G}_{32} \, d^4X_3 \, d^4X_4
\label{eqn2.14}
\end{equation}
where the usual rules of matrix multiplication are used.

\begin{figure}
\center{\includegraphics[width=12cm]{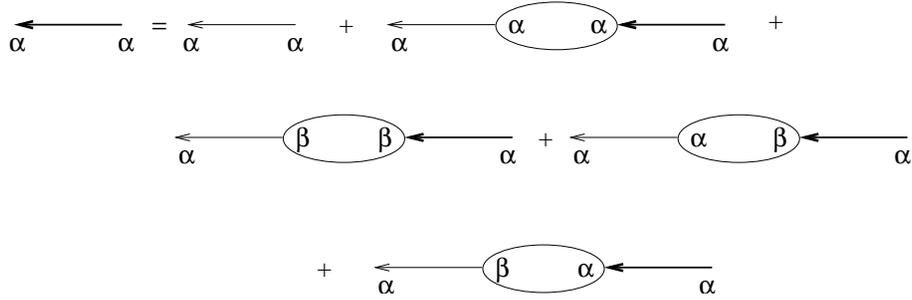}}
\caption{Dyson's equation in diagrammatic form for the Keldysh Green's function $G_{12}^{\alpha \alpha}$.  The thick line represents the exact Green's function $G_{12}^{\alpha \alpha}$, the thin line represents the Green's function for an ideal gas $G_{12}^{(0)\alpha \alpha}$, and the ellipse represents the self-energy $\Sigma$.}
\label{fig3}
\end{figure}

As we saw earlier, the components of the Green's function matrix are not independent, but are linearly related.  One can therefore perform a transformation to set one of the matrix components to zero.  There are many ways to do this; the one we shall use here is the one employed in most recent literature on the quasiclassical theory of superconductivity.\cite{rammer}  The matrices above are first rotated in Keldysh space using the transformation $\check{G} \rightarrow \tau_3 \check{G}$,
where the $\tau$ matrices are identical in form to the Pauli spin matrices
\begin{equation}
\tau^0=
\begin{pmatrix}
1 & 0 \\
0 & 1
\end{pmatrix},
\quad
\tau^1=
\begin{pmatrix}
0 & 1 \\
1 & 0
\end{pmatrix},
\quad
\tau^2=
\begin{pmatrix}
0 & -i \\
i & 0
\end{pmatrix},
\quad
\tau^3=
\begin{pmatrix}
1 & 0 \\
0 & -1
\end{pmatrix},
\label{eqn2.15}
\end{equation}
and then transformed $\check{G} \rightarrow Q \check{G} Q^\dagger$, where the matrix $Q$ is given by 
\begin{equation}
Q=\frac{1}{\sqrt{2}}(\tau^0 -i \tau^2).
\label{eqn2.16}
\end{equation}
The resulting matrices have the form
\begin{equation}
\check{G}=
\begin{pmatrix}
G^{R} &   G^{K}  \\
0 &   G^{A}
\end{pmatrix}, 
\qquad
\check{\Sigma}=
\begin{pmatrix}
\Sigma^{R} &   \Sigma^{K}  \\
0 &   \Sigma^{A}
\end{pmatrix}
\label{eqn2.17}
\end{equation}
where the retarded and advanced Green's functions $G^R$ and $G^A$ have been defined above, and the new Keldysh Green's function $G^K$ is given by 
\begin{equation}
G^K=G^{\alpha \alpha} + G^{\beta \beta} = G^{\alpha \beta} + G^{\beta \alpha}.
\label{eqn2.18}
\end{equation}
$\Sigma^R$, $\Sigma^A$ and $\Sigma^K$ are defined in a similar manner.  In what follows, we shall use the `check' to denote the Keldysh matrices.  In order to avoid writing integrals over the space and time coordinates, we introduce the binary operator $\otimes$, which has the effect of integrating over the free space and time coordinates and performing matrix multiplication when applied between two Keldysh matrices.  Thus, the Dyson equation for the Keldysh Green's function written above can be represented as
\begin{equation}
\check{G}=\check{G}^0 + \check{G}^0 \otimes \check{\Sigma} \otimes \check{G}
\label{eqn2.19}
\end{equation}
  
$\check{G}^0$ obeys the equation of motion
\begin{equation}
G_{01}^{-1}  \check{G}_{12}^0 =  \delta(X_1 - X_2).
\label{eqn2.20}
\end{equation}
Operating $G_{01}^{-1} $ on the Dyson equation Eqn(\ref{eqn2.19}) from the left, we obtain
\begin{equation}
G_{01}^{-1}\check{G}=\delta(X_1 - X_2) +  \check{\Sigma} \otimes \check{G}.
\label{eqn2.21}
\end{equation}
The conjugate equation is\footnote{From the right, we operate on the coordinate 2, hence we use the operator $G_{02}^{-1}$.  $G_{02}^{-1}$ operated on the right is equivalent to $G_{02}^{-1*}$ operated on the left.}  
\begin{equation}
\check{G}G_{02}^{-1}=\delta(X_1 - X_2) +   \check{G} \otimes \check{\Sigma} .
\label{eqn2.22}
\end{equation}
Subtracting the two equations, we obtain
\begin{equation}
(G_0^{-1} \otimes \check{G} -  \check{G} \otimes G_0^{-1}) - (\check{\Sigma} \otimes \check{G} - \check{G} \otimes \check{\Sigma})  =0,
\label{2.23}
\end{equation}
where we have suppressed the subscript `1' of $G_{01}^{-1}$.  This can be written as a commutator
\begin{align}
[G_0^{-1} \ominus \check{G}] - [\check{\Sigma} \ominus \check{G}] &=0 \nonumber \\
\intertext{or} 
[G_0^{-1} \ominus \check{G}] &=[\check{\Sigma} \ominus \check{G}]
\label{eqn2.24}
\end{align}
where the $\ominus$ operator defines the commutator of two operators $[A \ominus B] = A \otimes B - B \otimes A$. \\

\noindent \textit{The mixed or Wigner representation}

The Green's function $\check{G}_{12}$ oscillates rapidly with the difference $\vec{r}_2$ -  $\vec{r}_1$, on the scale of the inverse Fermi wave vector $k_F^{-1}$.  In the physical systems to be discussed, we are interested in variations on much longer length scales.  We therefore perform a transformation to center-of-mass coordinates, $\vec{R}$, $T$, and difference coordinates, $\vec{r}$, $\vec{t}$, defined by the equations \cite{kadanoff}
\begin{subequations} \label{eqn2.25}
\begin{align}
\vec{r}_1 =\vec{R}-\vec{r}/2,  &\qquad \vec{r}_2=\vec{R}+\vec{r}/2 \label{eqn2.25a} \\ 
t_1=T-t/2,    &\qquad t_2=T+t/2   \label{eqn2.25b}
\end{align}
\end{subequations}
and then define a Fourier transform of $\check{G}_{12}$ with respect to the variables $\vec{r}$ and $t$:
\begin{equation}
\check{G}(\vec{R}, T; \vec{r}, t)= \int e^{-i E t} e^{i \vec{p} \cdot \vec{r}} \; \check{G}(\vec{R}, T; \vec{p}, E)\; d\vec{p} \;d E.
\label{eqn2.26}
\end{equation}
In Eqn.(\ref{eqn2.24}), the part of the commutator involving $G_{0}^{-1}$ can be written as
\begin{equation}
 G_{01}^{-1} \otimes  \check{G} -  \check{G} \otimes  G_{02}^{-1} =  
\left[ i \left(\frac{\partial}{\partial t_1} + \frac{\partial}{\partial t_2}\right) - \frac{1}{2m}(\nabla_1^2-\nabla_2^2)\right] \check{G}
 \label{eqn2.27}
 \end{equation}
 using Eqn.(\ref{eqn2.10}) in the free electron approximation.  Transforming to the coordinates, $\vec{R}, T, \vec{r}, t$, this can be written
\begin{equation}
\left( i \frac{\partial}{\partial T} + \frac{1}{m}\nabla_{\vec{r}} \cdot \nabla_{\vec{R}} \right) \check{G}
\label{eqn2.28}
\end{equation}
From Eqn.(\ref{eqn2.18}), the Keldysh component $G^K$ of the Green's function can be written as the sum of the Green's functions $G^{\alpha \beta}$ and $G^{\beta \alpha}$.  Keeping in mind the definitions of these functions given in Eqns(\ref{eqn2.3}), we can define a non-equilibrium distribution function $f(\vec{R},T,\vec{r})$, which is related to the function  $G^{\beta \alpha}$ by
\begin{equation}
f(\vec{R},T,\vec{r})= \left< \hat{\psi}^+(\vec{R}+\vec{r}/2, T+t/2) \hat{\psi}(\vec{R}-\vec{r}/2, T-t/2 \right> _{t=0}= -i G^{\beta \alpha}(\vec{R}, T, \vec{r}, t)_{t=0}.
\label{eqn2.29}
\end{equation}
Similarly
\begin{equation}
 \left< \hat{\psi}(\vec{R}+\vec{r}/2, T+t/2) \hat{\psi}^+(\vec{R}-\vec{r}/2, T-t/2 \right> _{t=0}=1-f(\vec{R},T,\vec{r})  = i G^{\alpha \beta}(\vec{R}, T, \vec{r}, t)_{t=0}.
\label{eqn2.30}
\end{equation} 
Subtracting the first equation from the second, we obtain
\begin{equation}
1-2 f(\vec{R},T,\vec{r}) =  h(\vec{R},T,\vec{r})=i (G^{\alpha \beta}(\vec{R}, T, \vec{r}, t)_{t=0} + G^{\beta \alpha}(\vec{R}, T, \vec{r}, t)_{t=0} ) = i G^K(\vec{R}, T, \vec{r}, t)_{t=0},
\label{eqn2.31}
\end{equation}
where we have defined a new distribution function $h(\vec{R},T,\vec{r})$.  In terms of the mixed Fourier transform, Eqn.(\ref{eqn2.26}), this can be written as
\begin{equation}
h(\vec{R},T,\vec{r})=i G^K(\vec{R}, T, \vec{r}, t)_{t=0}=\frac{i}{2 \pi} \int e^{i \vec{p} \cdot \vec{r}} \; G^K(\vec{R}, T; \vec{p}, E)\; d\vec{p} \;d E.
\label{eqn2.32}
\end{equation}
Taking the Keldysh component of Eqn.(\ref{eqn2.27}) at $t=0$, we obtain
\begin{equation}
\left( i \frac{\partial}{\partial T} + \frac{1}{m}\nabla_{\vec{r}} \cdot \nabla_{\vec{R}} \right)  \int d\vec{p} \; \frac{d E}{2 \pi} \;e^{i \vec{p} \cdot \vec{r}} \; G^K(\vec{R}, T; \vec{p}, E) ,
\label{eqn2.33}
\end{equation}
or, in terms of the Fourier components with respect to $\vec{r}$, corresponding to momentum $\vec{p}$,
\begin{equation}
\frac{\partial h(\vec{R},T,\vec{p}) }{\partial T} + \frac{\vec{p}}{m} \cdot \nabla_{\vec{R}}h(\vec{R},T,\vec{p})
\label{eqn2.34}
\end{equation}
where the Fourier transform of the Wigner distribution function $h(\vec{R},T,\vec{p})$ is given by
\begin{equation}
h(\vec{R},T,\vec{p}) = \frac{i}{2 \pi}  \int  G^K(\vec{R}, T; \vec{p}, E) \;d E.
\label{eqn2.35}
\end{equation}
In equilibrium, $h$ is given by
\begin{equation}
h_0(\epsilon_{\vec{p}})=1-2 f_0(\epsilon_{\vec{p}})=\tanh(\epsilon_{\vec{p}}/2k_B T).
\label{eqn2.36}
\end{equation}
Eqn(\ref{eqn2.34}), which is the Keldysh component of the left hand side of Eqn(\ref{eqn2.24}), has the form of the one side of the classical Boltzmann equation for the distribution function.  (Using the definition of the function $h(\vec{R},T,\vec{p})$, this can also be written in a more conventional form in terms of $f(\vec{R},T,\vec{p})$.)  The right hand side of the Keldysh component of Eqn.(\ref{eqn2.24}) must therefore correspond to the collision terms.  The right hand side of the Keldysh component can be written as
$2 (\Sigma^{\beta \alpha} G^{\alpha \beta} - \Sigma^{\alpha \beta} G^{\beta \alpha})$.  Taking the limit at $t=0$, and writing in terms of the distribution function $f(\vec{R},T,\vec{p})$ using Eqn.(\ref{eqn2.28}) and Eqn.(\ref{eqn2.29}), this Keldysh component of the right hand side of Eqn(\ref{eqn2.24}) can be written as \cite{lifshitz}
\begin{equation}
-2 i [\Sigma^{\beta \alpha}(R,T,p)(1-f(\vec{R},T,\vec{p})) + \Sigma^{\alpha \beta}(R,T,p)f(\vec{R},T,\vec{p})].
\label{eqn2.37}
\end{equation}
The first term in Eqn.(\ref{eqn2.37}) with the factor $(1-f(\vec{R},T,\vec{p}))$ has the usual form of a scattering-in term, corresponding to the gain of particles, while the second term has the form for a scattering-out term, corresponding to a loss of particles.  Consequently, we see that the Keldysh component of the right-left subtracted Dyson equation gives the transport equation for the distribution function.  From the diagonal components components of the same equation, one can obtain solutions for the other components of $\check{G}$ and $\check{\Sigma}$.
More typically, the scattering terms on the right hand side of the Boltzmann equation make it difficult to solve, and some approximations must be employed.  If the variation of the system with the center-of-mass coordinates $T$ and $\vec{R}$ is small, then one can expand the Green's functions and self-energies, which are functions of $\vec{R}, \vec{r}, T, t$ about $\vec{R}$ and $T$ in a Taylor's series to first order in $\vec{r}$ and $t$.  This is the gradient expansion discussed by Kadanoff and Baym\cite{kadanoff} and Larkin and Ovchinnikov\cite{larkin}, and we shall return to it at the end of this section.  

Instead of taking the difference of Eqn(\ref{eqn2.21}) and its conjugate equation, Eqn(\ref{eqn2.22}), we take the sum, we obtain the equation
\begin{equation}
[G_0^{-1} \oplus \check{G}] =2 \delta(X_1-X_2) + [\check{\Sigma} \oplus \check{G}]
\label{eqn2.38}
\end{equation}
where the operator $\oplus$ defines the Keldysh anti-commutator, in the same way as the operator $\ominus$ defines the Keldysh commutator.  The left hand side of Eqn.(\ref{eqn2.38}) can be written as
\begin{equation}
 G_{01}^{-1} \otimes  \check{G} +  \check{G} \otimes  G_{02}^{-1} =  
\left[ i \left(\frac{\partial}{\partial t_1} - \frac{\partial}{\partial t_2}\right) - \frac{1}{2m}(\nabla_1^2+\nabla_2^2)\right] \check{G}.
\label{eqn2.39}
\end{equation}
Transforming to the mixed representation, we obtain
\begin{equation}
\left[  2 i \frac{\partial}{\partial t} + \frac{1}{m} \left( \frac{1}{4}\nabla_{\vec{R}}^2 + \nabla_{\vec{r}}^2 \right) 
\right] \check{G}(\vec{R}, T, \vec{r}, t)
\label{eqn2.40}
\end{equation}
Now, the assumption we are making is that the variations of $G$ on the scale of $\vec{R}$ are much slower than the variations on the scale of $\vec{r}$.  Hence, the the terms in equation above involving derivatives with $\vec{R}$ contribute much less than the those with $\vec{r}$, and can be neglected in this approximation.  If we consider the equation for $G_0$, for which the terms involving $\Sigma$ on the right hand side of Eqn.(\ref{eqn2.38}) are 0, then we obtain, after transforming to Fourier components\cite{ketterson}
\begin{align}
&(E-\epsilon_{\vec{p}})G_0(\vec{R}, T, \vec{p}, E) =1 \label{eqn2.41}\\
\intertext{or}
&G_0(\vec{R}, T, \vec{p}, E) = \frac{1}{(E-\epsilon_{\vec{p}})}, \label{eqn2.42}
\intertext{where}
&\epsilon_{\vec{p}} = \frac{p^2}{2 m} - \mu \sim v_F(p-p_F). \label{eqn2.43}
\end{align}
So far, we have assumed a free-electron model.  If there is a slowly varying potential $U(\vec{R}, T)$, the equation above can be modified to 
\begin{equation}
G_0(\vec{R}, T, \vec{p}, E)= \frac{1}{(E-\epsilon_{\vec{p}}-U(\vec{R},T))}
\label{eqn2.44}
\end{equation}
This equation has form of a Green's function for a free particle, but in a slowly varying potential $U(\vec{R}, T)$.  The operator $G_0^{-1}$ in the mixed representation can therefore be written as
\begin{equation}
G_0^{-1} =   (E-\epsilon_{\vec{p}}-U(\vec{R},T)).
\label{eqn2.45}
\end{equation}
More accurately, one transforms Eqn.(\ref{eqn2.10}) using the following equations
\begin{subequations} \label{eqn2.46}
\begin{align}
\frac{\partial}{\partial t_{1,2} }&= \frac{1}{2} \frac{\partial}{\partial T} \pm \frac{\partial}{\partial t}, \label{eqn2.46a} \\
\nabla_{1,2} &=\frac{1}{2}\nabla_{\vec{R}} \pm \nabla_{\vec{r}} \label{eqn2.46b} \\
\nabla^2_{1,2} &= \frac{1}{4} \nabla^2_{\vec{R}} \pm \nabla_{\vec{r}} \cdot \nabla_{\vec{R}} + \nabla^2_{\vec{r}}.
\label{eqn2.46c} 
\end{align}
\end{subequations}
Again, assuming the functions in the mixed representation are slowly varying functions of $\vec{R}$, we can ignore the second derivatives with respect to $\vec{R}$, to obtain
\begin{equation}
G_0^{-1}=\frac{1}{2} \frac{\partial}{\partial T} + \frac{\partial}{\partial t}  + \frac{\nabla^2_{\vec{r}}}{2m}+ \frac{\nabla_{\vec{r}} \cdot \nabla_{\vec{R}}}{2m} + \mu
\label{eqn2.47}
\end{equation}
In most applications of interest here, we also ignore the slow $T$ dependence.  Adding the potential $U(\vec{R},T)$, and writing in terms of the partial Fourier transform with respect to $\vec{p}$ and $E$, we obtain
\begin{equation}
G_0^{-1}=  E- \epsilon_{\vec{p}} + \frac{i}{2} \vec{v}_F \cdot \nabla_{\vec{R}} + \mu - U(\vec{R}, T),
\label{eqn2.48}
\end{equation}
where we have replaced $\vec{p}/m$ by $\vec{v}_F$, since the important region of interest is at the Fermi surface.  Note that we still keep account of the \textit{direction} of the momentum.
 
To conclude this section, we derive some expressions for the convolution of two operators in the mixed representation, the gradient expansion discussed above.  Consider the convolution of two operators defined as an integral over the internal space and time coordinates
\begin{equation}
A \otimes B = \int d\vec{r}_3 \int dt_3 A(\vec{r}_1, \vec{r}_3, t_1,t_3) B(\vec{r}_3, \vec{r}_2, t_3, t_2)
\label{2.49}
\end{equation}
If we do a partial transform to $\vec{p}, \vec{R}$ (but do not transform the time coordinates), $A \otimes B$ can be represented as
\begin{equation}
A \otimes B = \int dt_3 e^{\frac{i}{2} \left( \nabla^A_{\vec{R}} \cdot \nabla^B_{\vec{p}} -  \nabla^A_{\vec{p}} \cdot \nabla^B_{\vec{R}} \right)} A(\vec{R}, \vec{p}, t_1, t_3) B(\vec{R}, \vec{p}, t_3, t_2) 
\label{eqn2.50}
\end{equation}
where the superscripts to the derivatives denote that they operate only those functions.  If the derivatives are small, we need to take only the first order expansion of this expression
\begin{equation}
(A \otimes B) (\vec{p}, \vec{R})= \int dt_3 \left[1+ \frac{i}{2} \left( \nabla^A_{\vec{R}} \cdot \nabla^B_{\vec{p}} -  \nabla^A_{\vec{p}} \cdot \nabla^B_{\vec{R}} \right) \right] A(\vec{R}, \vec{p}, t_1, t_3) B(\vec{R}, \vec{p}, t_3, t_2). 
\label{eqn2.51}
\end{equation}
Similarly
\begin{equation}
(B \otimes A) (\vec{p}, \vec{R})= \int dt_3 \left[1+ \frac{i}{2} \left( \nabla^B_{\vec{R}} \cdot \nabla^A_{\vec{p}} -  \nabla^B_{\vec{p}} \cdot \nabla^A_{\vec{R}} \right) \right] B(\vec{R}, \vec{p}, t_1, t_3) A(\vec{R}, \vec{p}, t_3, t_2).
\label{eqn2.52}
\end{equation}
If we were dealing with functions alone, then the multiplication of the two function $A$ and $B$ is commutative.  When $A$ and $B$ are matrices, however, they do not in general commute, so we obtain
\begin{equation}
(A \ominus B) (\vec{p}, \vec{R}) = A \otimes B - B \otimes A= \int dt_3 \; [A,B] + 
\frac{i}{2} \int dt_3 \; \left[ \{\nabla_{\vec{R}}A, \nabla_{\vec{p}}B\} - \{\nabla_{\vec{p}}A, \nabla_{\vec{R}} B \} \right],
\label{eqn2.53}
\end{equation}
where $[A,B]$ notation stands for the commutator of the two functions, and $\{A,B\}$ stands for the anticommutator.
A similar equation can be obtained if we transform the times to the mixed representation $T, E$
\begin{equation}
(A \ominus B) (T, E) = A \otimes B - B \otimes A= \int d\vec{r}_3 \; [A,B] + 
\frac{i}{2} \int d\vec{r}_3 \; \left[ \{\partial_{T}A, \partial_{E}B\} - \{\partial_{E}A, \partial_{T} B \} \right],
\label{eqn2.54}
\end{equation}
When we transform both sets of variables, we obtain
\begin{equation}
\begin{split}
(A \ominus B) (\vec{p}, \vec{R}, T, E) &=  A \otimes B - B \otimes A \\
&= [A,B] + 
\frac{i}{2}\left[\left( \{\partial_{T}A, \partial_{E}B\} - \{\partial_{E}A, \partial_{T} B \}\right)+ \left(\{\nabla_{\vec{R}}A, \nabla_{\vec{p}}B\} - \{\nabla_{\vec{p}}A, \nabla_{\vec{R}} B \} \right) \right].
\end{split}
\label{eqn2.55}
\end{equation}
In most cases, we are interested in stationary situations, where there is no $T$ dependence.  In this case, the equation above reduces to
\begin{equation}
(A \ominus B) (\vec{p}, \vec{R}, T, E) = A \otimes B - B \otimes A= [A,B] + 
\frac{i}{2}\left[ \{\nabla_{\vec{R}}A, \nabla_{\vec{p}}B\} - \{\nabla_{\vec{p}}A, \nabla_{\vec{R}} B \}  \right].
\label{eqn2.56}
\end{equation}
These expressions will be useful in the calculations to follow.

\section{The quasiclassical approximation}
The non-equilibrium spectral function $A$ is defined in the same way as in the equilibrium case\cite{rammer,abrikosov}
\begin{equation}
A= \frac{i}{2 \pi} (G^R - G^A)=-\frac{1}{\pi} \Im (G^R),
\label{eqn3.1}
\end{equation}
where $\Im (G^R)$ denotes the imaginary part of the retarded Green's function.  In the equilibrium case, $A$ defines the spectrum of energy levels, for stationary quantum states, it has the form of a sum of $\delta$-functions at each state energy.  In the quasiparticle approximation, these $\delta$-functions are broadened, but the width $\Gamma$ of each state, defining its lifetime, is still small compared to its energy.  If $\Gamma$ is large, the quasiparticle approximation breaks down, and one cannot obtain a kinetic equation for a distribution function by integrating over the energy $E$.  However, for most perturbations of interest, the self-energies typically have a weak dependence on the magnitude of the momentum, this dependence being appreciable only near the Fermi energy.  On the other hand, while the Dyson equation has a strong dependence on $E$ and $\epsilon_{\vec{p}}$, the subtracted Dyson equation, Eqn.(\ref{eqn2.24}), from which we will obtain the equations of motion, has a very weak dependence on both $E$ and $\epsilon_{\vec{p}}$.  In this case, it is possible to average over the particle energy to eliminate the dependence on the magnitude of the momentum, but keep the dependence on the direction of the momentum.  Hence, one can think of replacing the Green's functions and self-energies by their values on the Fermi surface, multiplied by a $\delta$-function in the form
\begin{equation}
G(\vec{R}, \vec{p}, t_1, t_2) \rightarrow \delta(\epsilon_{\vec{p}}) g(\vec{R}, \hat{p}, t_1, t_2)
\label{eqn3.2}
\end{equation} 
To this end, we define the so-called quasiclassical Green's function
\begin{equation}
g( \vec{R}, \hat{p}, t_1, t_2)=  \frac{i}{\pi} \int d \epsilon_{\vec{p}} \; G(\vec{R}, \vec{p}, t_1, t_2).
\label{eqn3.3}
\end{equation}
Care must be taken in performing this integral, since the integrand falls off only as $1/ \epsilon_{\vec{p}}$ for large $ \epsilon_{\vec{p}}$.  To avoid this, one can introduce a cut-off in the integral, as done by Serene and Rainer,\cite{serene} or following Eilenberger,\cite{eilenberger} use a special contour for integration.

We would like to obtain an equation of motion for the quasiclassical Green's functions.  If we obtain an equation of motion by operating $G_0^{-1}$ in the form of Eqn(\ref{eqn2.47}) on Dyson's equation, and then integrating over $\epsilon_{\vec{p}}$, there are terms in the equation that will have large contributions.  In order to eliminate these large terms, we start from the left-right subtracted equation of motion, Eqn.(\ref{eqn2.24}), in which the troublesome terms are cancelled, and then integrate over $\epsilon_{\vec{p}}$.  The terms on the right hand side of Eqn.(\ref{eqn2.24}) are of the form
\begin{equation}
 \check{\Sigma} \otimes \check{G} = \int dt_3 \; dx_3^3 \; \check{\Sigma}(x_1, t_1, x_3, t_3)  \check{G}(x_3, t_3, x_2, t_2) 
\label{eqn3.4}
 \end{equation}
We first Fourier transform this term with respect to $\vec{p}$, which takes care of the integral over $x_3$.  We then average the resulting equation with respect to $\epsilon_{\vec{p}}$.  The assumption here is that only $\check{G}$ has a strong dependence on the momentum $\vec{p}$, so the result of this averaging is a term of the form
\begin{equation}
\int dt_3  \; \check{\Sigma}(\vec{R}, \vec{p}, t_1, t_3) \check{g}( \vec{R}, \hat{p}, t_3, t_2)
\label{eqn3.5}
\end{equation}
Now, to complete the transformation, the self-energy $\check{\Sigma}$, which is a functional of the Green's functions $\check{G}$, must become a functional only of the quasiclassical Green's functions $\check{g}$, $\check{\Sigma}[\check{G}] \rightarrow \check{\sigma}[\check{g}]$.  With this final change, Eqn.(\ref{eqn2.24}) for the quasiclassical Green's functions becomes
\begin{equation}
[(g_0^{-1} -\check{\sigma} ) \underset {^,}{\circ} \check{g}]=0
\label{eqn3.6}
\end{equation} 
The $\circ$ operator in the commutator $[A \underset {^,}{\circ} B]$ involves an integral over the internal time coordinates in addition to the usual matrix multiplication for Keldysh matrices.  If we transform the time coordinates as well, this integral can be removed, as shown at the end of the last section (Eqn.{\ref{eqn2.55}).  In this case, the commutator becomes a simple commutator (but involving matrix multiplication of the Keldysh matrices).  

It remains to express the physical quantities of interest in terms of the Green's functions.  The particle density is given by 
\begin{equation}
\rho(1)=-2 i G_{11}^{\beta \alpha}
\label{eqn3.7}
\end{equation}
and the current density in the absence of external fields is given by
\begin{equation}
\vec{j}(1)= -\frac{e}{m} \left( \nabla_1 - \nabla_2 \right)G_{12}^{\beta \alpha}\arrowvert_{2=1}
\label{eqn3.8}
\end{equation}
From the definitions of $G^R$, $G^A$ and $G^K$  in terms of $G^{\alpha \beta}$, $G^{\beta \alpha}$ and $G^{\beta \beta}$, we can write the function $G^{\beta \alpha}$ as
\begin{equation}
G^{\beta \alpha}=\frac{1}{2} \left[G^K+(G^R - G^A)\right].
\label{eqn3.9}
\end{equation}
$G^R$ and $G^A$ depend on the equilibrium properties of the system, and so do not contribute to the current on nonequilibrium density.  Consequently, these terms can be dropped in the expression for the particle density and current.  Writing in terms of the quasiclassical Green's functions, we can obtain expressions for the density and the current in the mixed representation.
Consider the expression for charge density
\begin{equation}
\delta \rho(\vec{R},T) = - i e G^K(\vec{R},T, \vec{r}=0, t=0),
\label{eqn3.10}
\end{equation}
where we use $\delta \rho$ instead of $\rho$ to emphasize that this does not include the equilibrium contributions.  Expanding $G^K$ in terms of Fourier components in momentum space $\vec{p}$
\begin{align}
\delta \rho(\vec{R},T) & = - i e \left[ \int \frac{d^3p}{(2 \pi)^3} \; e^{i \vec{p}\cdot\vec{r}} G^K(\vec{R}, T, \vec{p}, t) \right]_{\vec{r}=0, t=0} \nonumber \\
&= - i e \int \frac{dE}{2 \pi} \; \int \frac{d^3p}{(2 \pi)^3} \; G^K(\vec{R}, T, \vec{p}, E) 
\label{eqn3.11} 
\end{align} 
where we have used the fact that setting $t=0$ in $G^K(t)$ is equivalent to integrating $G^K(E)$ over $dE/2 \pi$. 

The equivalent expression for the current density can also be written in terms of the quasiclassical Green's function.  From Eqn(\ref{eqn3.8}) and Eqn(\ref{eqn3.9}), we have, in the mixed representation
\begin{equation}
\vec{j}(\vec{R}, T)= -\frac{e}{2m}\left[ \left( \nabla_1 - \nabla_2 \right)G^K(\vec{R},T, \vec{r}, \vec{p} ) \right]_{\vec{r}=0, t=0}
\label{eqn3.12}
\end{equation}
The integral over the momentum $\vec{p}$ can be rewritten as
\begin{equation}
 \int \frac{d^3p}{(2 \pi)^3} \rightarrow N_0 \int d\epsilon_{\vec{p}} \int \frac{d \Omega_p}{4 \pi}
 \label{eqn3.13}
 \end{equation}
where $N_0$ is the density of states at the Fermi energy, and $d\Omega_p$ is an element of solid angle in momemtum space $p$.  Writing $\nabla_1 - \nabla_2 = 2 \nabla_{\vec{r}}$ from Eqn.(\ref{eqn2.46b}), we obtain
\begin{equation}
\vec{j}(\vec{R},T) = -\frac{e}{m} \left[ \nabla_{\vec{r}}  \int \frac{d^3p}{(2 \pi)^3} \; e^{i \vec{p}\cdot\vec{r}} G^K(\vec{R}, T, \vec{p}, t) \right]_{\vec{r}=0, t=0}.
\label{eqn3.14}
\end{equation}
Operating $\nabla_{\vec{r}}$ on the exponential within the integral gives $i \vec{p}$.  In the quasiclassical approximation, we assume that the major contribution comes from near the Fermi surface, so that $\vec{p}/m = \vec{v}_F$, the Fermi velocity.  With this approximation, we obtain
\begin{equation}
\vec{j}(\vec{R}, T)  = -i e \int \frac{dE}{2 \pi}  \int \frac{d^3p}{(2 \pi)^3} \; \vec{v}_F G^K(\vec{R}, T, \vec{p}, E). 
\label{eqn3.15}
\end{equation}
If we use the definition of the distribution function $h$ given by Eqn.(\ref{eqn2.31}), we obtain
\begin{equation}
\vec{j}(\vec{R}, T)  = -e  \int \frac{d^3p}{(2 \pi)^3} \; \vec{v}_F h(\vec{R}, T, \vec{p})
\label{eqn3.16}
\end{equation}
which is the expected classical form for the current.  Recalling the definition of the quasiclassical Green's function, Eqn(\ref{eqn3.3}), we obtain 
\begin{equation}
\vec{j}(\vec{R}, T) = -\frac{1}{2} e N_0 \int dE \int \frac{d\Omega_p}{4 \pi} \vec{v}_F g^K(E,\hat{p}, \vec{R}, T)
\label{eqn3.17}
\end{equation}
where we have used Eqn(\ref{eqn3.3}) and Eqn.(\ref{eqn3.13}).  Note that writing these expressions in terms of the quasiclassical Green's functions necessitates reversing the order of integration over $E$ and $\epsilon_{\vec{p}}$.

The transformation, Eqn.(\ref{eqn3.13}), assumes that the density of states is constant at the Fermi energy, and hence assumes that particle-hole symmetry holds.  Hence, within this approximation, we will not be able to obtain any results on physical phenomenon that depend on particle-hole asymmetry, in particular, thermoelectric effects.  This should be contrasted with the derivation of the current in the diffusive limit that we performed in the introduction, where the energy dependence of the density of states was taken into account explicitly.

So far, we have ignored the effects of external fields and potentials.  In particular, the effect of a magnetic field is of interest.  The magnetic field is introduced in the form of a vector potential $\vec{A}(\vec{R},T)$.  Here, we shall consider only time independent fields.  $\vec{A}(\vec{R})$ is introduced by making the change
\begin{equation}
\nabla_{\vec{R}} \rightarrow \nabla_{\vec{R}}-i e \vec{A}(\vec{R}) \equiv \partial_{\vec{R}}
\label{eqn3.18}
\end{equation}  
in all equations involving the spatial derivative.  For example, the operator $\tilde{G}_{01}^{-1}$ is now written
\begin{equation}
\tilde{G}_{01}^{-1} = i \frac{\partial}{\partial t_1} + \frac{\partial^2_{\vec{r}}}{2m} + \mu - e \phi.
\label{eqn3.19}
\end{equation}
In the equation above, we have also added a term $e \phi$, corresponding to the presence of a scalar potential.

In our formulation, we would like all observable quantities to be invariant under a gauge transformation of the vector potential
\begin{equation}
\vec{A} \rightarrow \vec{A} + \nabla \chi(\vec{r}),
\label{eqn3.20}
\end{equation}
which also transforms the potential
\begin{equation}
\phi \rightarrow \phi - \frac{\partial \chi}{\partial t}.
\label{eqn3.21}
\end{equation}
Eqn.(\ref{eqn3.14}) for the electrical current is then modified to
\begin{equation}
\vec{j}(\vec{R},T) = -\frac{2e}{m} \left[ [\nabla_{\vec{r}}-ie\vec{A}(\vec{R})]  \int \frac{d^3p}{(2 \pi)^3} \; e^{i \vec{p}\cdot\vec{r}} G^{\beta \alpha}(\vec{R}, T, \vec{p}, t) \right]_{\vec{r}=0, t=0}, 
\label{eqn3.22}
\end{equation}
where we have written the current in terms of $G^{\beta \alpha}$ rather than $G^K$.  It follows that Eqn.(\ref{eqn3.15}) can be written as
\begin{equation}
\vec{j}(\vec{R}, T)  = -i 2 e \int \frac{dE}{2 \pi}  \int \frac{d^3p}{(2 \pi)^3} \; \vec{v}_F G^{\beta \alpha} (\vec{R}, T, \vec{p}, E) + \frac{2ie^2}{m} \vec{A}(\vec{R}) \int \frac{dE}{2 \pi}  \int \frac{d^3p}{(2 \pi)^3} \;G^{\beta \alpha}(\vec{R}, T, \vec{p}, E). 
\label{eqn3.23}
\end{equation}
The second term in the equation above is called the diamagnetic term.  It is cancelled by a contribution from the first term arising from energies far from the Fermi energy, which are not taken into account in the quasiclassical Green's function as defined by Eqn.(\ref{eqn3.3}).\cite{rammer}  Note that such a cancellation does not occur in the case of a superconductor, and the second term gives rise to the supercurrent, which is proportional to the vector potential and the phase gradient according to Eqn.(\ref{eqn3.21}).  With this high energy contribution cancelled by the diamagnetic term, the equation above transforms into the expression Eqn.(\ref{eqn3.17}) for the electrical current, written now in terms of the Keldysh component of the quasiclassical Green's function.

Such a cancellation does not occur in transforming $\delta \rho$ in terms of the quasiclassical Green's functions, and the contribution of the integrals in Eqn.(\ref{eqn3.11}) must be explicitly calculated.\cite{rammer}  The result that was obtained by Eliashberg\cite{eliashberg} can be written as
\begin{equation}
\delta \rho(\vec{R},T) = -\frac{e N_0}{2} \int dE \int \frac{d\Omega_p}{4 \pi} g^K(E,\hat{p}, \vec{R}, T)-2e^2 N_0 \phi(\vec{R},T),
\label{eqn3.24}
\end{equation}
where $\phi(\vec{R},T)$ is now the scalar electrochemical potential.  This expression  can also be obtained by invoking gauge invariance arguments.  Since we must preserve charge neutrality in the system, $\delta \rho =0$, so that $\phi(\vec{R}, T)$ is given by
\begin{equation}
\phi(\vec{R},T)=-\frac{1}{4e} \int dE \int \frac{d\Omega_p}{4 \pi} g^K(E,\hat{p}, \vec{R}, T)
\label{eqn3.25}
\end{equation}

For completeness, we also write expressions for the thermal current and the density of states. The expression for the thermal current in terms of the Keldysh Green's function is given by
\begin{equation}
\vec{j}_{th}(\vec{R}, T)  = i \int \frac{dE}{2 \pi}  \int \frac{d^3p}{(2 \pi)^3} \; E\vec{v}_F G^K(\vec{R}, T, \vec{p}, E),
\label{eqn3.26}
\end{equation}
and the corresponding form in terms of the quasiclassical Green's functions
\begin{equation}
\vec{j}_{th}(\vec{R}, T) = \frac{1}{2} N_0 \int dE \int \frac{d\Omega_p}{4 \pi} E \vec{v}_F g^K(E,\hat{p}, \vec{R}, T).
\label{eqn3.27}
\end{equation}

As in the equilibrium case, the density of states is given directly by the spectral function $A$ defined in Eqn.(\ref{eqn3.1}), now expressed in the mixed representation
\begin{align}
N(E, \vec{R}, T)=  A(E, \vec{R}, T) &= -\frac{1}{\pi} \Im G^R(E, \vec{R}, T) \nonumber \\
&= -\frac{1}{\pi} \Im \int \frac{d^3p}{(2 \pi)^3} G^R(E, \vec{p}, \vec{R}, T). \label{eqn3.28}
\end{align}
Written in terms of the quasiclassical Green's function, this becomes
\begin{equation}
N(E, \vec{R}, T) = N_0 \Re \int \frac{d\Omega_p}{4 \pi} g^R(\vec{R}, E, T),
\label{eqn3.29}
\end{equation}
where the notations $\Re$ and $\Im$ stand for the real and imaginary components respectively.

\section{Non-equilibrium Green's functions for superconducting systems}
The formalism that we have developed for non-equilibrium Green's functions for normal systems carries over into the superconducting case, except that the Green's functions for the superconducting case are more complicated.  In order to show how the Green's functions are defined, we start by expressing the Green's functions in the superconducting case in terms of field operators in the Nambu-Gorkov formalism.  In this formalism, the field operators can be written as two-component column matrices
\begin{equation}
\hat{\Psi}_1=
\begin{pmatrix}
\hat{\psi}_{1\uparrow} \\
\hat{\psi}^+_{1\downarrow}
\end{pmatrix}
\label{eqn4.1}
\end{equation}
where the up and down arrows refer to the spin indices, and the numbers now refer only to the time and space coordinates.  The Hermitian adjoint of this operator can be written as a two-component row matrix
\begin{equation}
\hat{\Psi}_1^+=
\begin{pmatrix}
\hat{\psi}^+_{1\uparrow} & \hat{\psi}_{1\downarrow}
\end{pmatrix}.
\label{eqn4.2}
\end{equation}
The multiplication of two such operators is defined as the tensorial product of the two matrices.  For example
\begin{equation}
\hat{\Psi}_1 \hat{\Psi}^+_2 =
\begin{pmatrix}
\hat{\psi}_{1\uparrow} \hat{\psi}_{2\uparrow}^+  &  \hat{\psi}_{1\uparrow} \hat{\psi}_{2\downarrow} \\
\hat{\psi}_{1\downarrow}^+ \hat{\psi}_{2\uparrow}^+  &  \hat{\psi}_{1\downarrow}^+ \hat{\psi}_{2\downarrow} 
\end{pmatrix}
\label{eqn4.3}
\end{equation}
It is natural then to define the Green's functions in terms of these products.  For example, the natural definition of  the Green's function corresponding to $G^{\alpha \alpha}$, Eqn.(\ref{eqn2.3a}), would be
\begin{equation}
G_{12}^{\alpha \alpha}=-i \left<T \hat{\Psi}_1 \hat{\Psi}_2^+\right> = - i 
\begin{pmatrix}
\left<T\hat{\psi}_{1\uparrow} \hat{\psi}_{2\uparrow}^+\right>  & \left<T \hat{\psi}_{1\uparrow} \hat{\psi}_{2\downarrow} \right>\\
\left<T\hat{\psi}_{1\downarrow}^+ \hat{\psi}_{2\uparrow}^+\right>  & \left<T\hat{\psi}_{1\downarrow}^+ \hat{\psi}_{2\downarrow}\right> 
\end{pmatrix}.
\label{eqn4.4}
\end{equation}
However, we would like to keep the same form for the equations of motion and Dyson's equation as for the normal case, except, of course, the quantities will be matrices in Nambu-Gorkov space.  To see if the definition above fits this requirement, let us operate $i\partial/\partial t_1$ on the Green's function for an ideal gas, as defined above.  With the definition
\begin{equation}
h_{01}=-\frac{\nabla_1^2}{2m} - \mu,
\label{eqn4.5}
\end{equation}
we have
\begin{equation}
i\frac{\partial}{\partial t_1} G_{12}^{(0)\alpha \alpha} = 
h_{01}
\begin{pmatrix}
\left<T\hat{\psi}_{1\uparrow} \hat{\psi}_{2\uparrow}^+\right>  & \left<T \hat{\psi}_{1\uparrow} \hat{\psi}_{2\downarrow} \right>\\
-\left<T\hat{\psi}_{1\downarrow}^+ \hat{\psi}_{2\uparrow}^+\right>  & -\left<T\hat{\psi}_{1\downarrow}^+ \hat{\psi}_{2\downarrow}\right> 
\end{pmatrix}
+
\begin{pmatrix}
\delta(X_1-X_2)  & 0 \\
0 & \delta(X_1-X_2) 
\end{pmatrix}
\label{eqn4.6}
\end{equation}
This does not have the required form of the analogous equation for the normal case, Eqn.(\ref{eqn2.9}).  To remedy this, we define our Green's functions with an extra factor of $\tau^3$.  The definitions corresponding to Eqns.(\ref{eqn2.3}) are then
\begin{subequations} \label{eqn4.7}
\begin{align}
\tilde{G}_{12}^{\alpha \alpha} & =-i \tau^3 \left<T \hat{\Psi}_1 \hat{\Psi}_2^+\right> = - i  \tau^3
\begin{pmatrix}
\left<T\hat{\psi}_{1\uparrow} \hat{\psi}_{2\uparrow}^+\right>  & \left<T \hat{\psi}_{1\uparrow} \hat{\psi}_{2\downarrow} \right>\\
\left<T\hat{\psi}_{1\downarrow}^+ \hat{\psi}_{2\uparrow}^+\right>  & \left<T\hat{\psi}_{1\downarrow}^+ \hat{\psi}_{2\downarrow}\right> 
\end{pmatrix}, \label{eqn4.7a} \\
\tilde{G}_{12}^{\beta \beta} & =-i \tau^3 \left<\tilde{T} \hat{\Psi}_1 \hat{\Psi}_2^+\right> = - i  \tau^3
\begin{pmatrix}
\left<\tilde{T}\hat{\psi}_{1\uparrow} \hat{\psi}_{2\uparrow}^+\right>  & \left<\tilde{T} \hat{\psi}_{1\uparrow} \hat{\psi}_{2\downarrow} \right>\\
\left<\tilde{T}\hat{\psi}_{1\downarrow}^+ \hat{\psi}_{2\uparrow}^+\right>  & \left<\tilde{T}\hat{\psi}_{1\downarrow}^+ \hat{\psi}_{2\downarrow}\right> 
\end{pmatrix}, \label{eqn4.7b} \\
\tilde{G}_{12}^{\alpha \beta} & =-i \tau^3 \left<\hat{\Psi}_1 \hat{\Psi}_2^+\right> = - i  \tau^3
\begin{pmatrix}
\left<\hat{\psi}_{1\uparrow} \hat{\psi}_{2\uparrow}^+\right>  & \left< \hat{\psi}_{1\uparrow} \hat{\psi}_{2\downarrow} \right>\\
\left<\hat{\psi}_{1\downarrow}^+ \hat{\psi}_{2\uparrow}^+\right>  & \left<\hat{\psi}_{1\downarrow}^+ \hat{\psi}_{2\downarrow}\right> 
\end{pmatrix}, \label{eqn4.7c} \\
\intertext{and}
\tilde{G}_{12}^{\beta \alpha} & =i \tau^3 \left<\hat{\Psi}_2^+ \hat{\Psi}_1\right> = i  \tau^3
\begin{pmatrix}
\left<\hat{\psi}_{2\uparrow}^+ \hat{\psi}_{1\uparrow}\right>  & \left<\hat{\psi}_{2\downarrow} \hat{\psi}_{1\uparrow} \right>\\
\left<\hat{\psi}_{2\uparrow}^+ \hat{\psi}_{1\downarrow}^+\right>  & \left<\hat{\psi}_{2\downarrow} \hat{\psi}_{1\downarrow}^+\right> 
\end{pmatrix}, \label{eqn4.7d}
\end{align}
\end{subequations}
where the `tilde' over the Green's functions denotes that they are matrices in Nambu-Gorkov space.
The operator corresponding to Eqn.(\ref{eqn2.9}) is then defined as
\begin{equation}
\tilde{G}_{01}^{-1} = i \tau^3 \frac{\partial}{\partial t_1} + \frac{\partial^2_1}{2m} + \mu.
\label{eqn4.8}
\end{equation}
where it is understood that any `scalar' quantities in the equation above and in what follows are mulitiplied by the identity matrix $\tau^0$.
With these modifications, all the equations derived for the Keldysh Green's functions for the  normal case can be carried over directly to the superconducting case.  In particular, the advanced and retarded Green's functions $\tilde{G}^R$ and $\tilde{G}^A$ are defined in terms of the Green's functions in Eqn.(\ref{eqn4.7}) in the same manner as before.  The Keldysh matrices corresponding to Eqn.(\ref{eqn2.17}) are then
\begin{equation}
\hat{G}=
\begin{pmatrix}
\tilde{G}^{R} &   \tilde{G}^{K}  \\
0 &   \tilde{G}^{A}
\end{pmatrix}, 
\qquad
\hat{\Sigma}=
\begin{pmatrix}
\tilde{\Sigma}^{R} &  \tilde{\Sigma}^{K}  \\
0 &   \tilde{\Sigma}^{A}
\end{pmatrix}.
\label{eqn4.9}
\end{equation}
Since each element of these matrices are themselves 2x2 matrices, the resulting Keldysh matrices for superconductors are 4x4 `supermatrices,' and we shall denote them by a `hat' symbol ( $\hat{}$ ).  The equation of motion equivalent to Eqn.(\ref{eqn2.24}) is
\begin{equation}
[\hat{G}_0^{-1} \ominus \hat{G}]=[\hat{\Sigma} \ominus \hat{G}].
\label{eqn4.10}
\end{equation}
Before we go any further, we need to specify the self-energy $\tilde{\Sigma}$.  At the temperatures of interest in the experiments on proximity systems, the important self-energy terms are due to electron-phonon scattering ($\tilde{\Sigma}_{e-p}$), and electron impurity scattering ($\tilde{\Sigma}_{imp}$). For conventional superconductors, the elastic component ($\tilde{\Sigma}^R_{e-p}+\tilde{\Sigma}^A_{e-p}$) of the electron-phonon contribution is the one that leads to the coupling between superconducting electrons\cite{schoen} (the other contributions of electron-phonon scattering will be ignored, under the assumption that we are at low enough temperatures so that inelastic electron-phonon scattering can be ignored).  Perhaps a simpler way of dealing with this component of the electron-phonon interaction is to start directly with the Gorkov equations of motion\cite{abrikosov} for the Green's function Eqn.(\ref{eqn4.7a}), which we write in the form (ignoring impurity scattering for the moment)
\begin{equation}
\begin{pmatrix}
\ i \frac{\partial}{\partial t_1} + \frac{\nabla^2_1}{2m} + \mu & \Delta \\
-\Delta^* &  -i \frac{\partial}{\partial t_1} + \frac{\nabla^2_1}{2m} + \mu
\end{pmatrix}
\tilde{G}_{12}^{\alpha \alpha} = \delta(X_1 - X_2),
\label{eqn4.11}
\end{equation}
with the pair potential $\Delta$ defined as\footnote{$\Delta$ defined this way differs from the conventional definition by a factor of $i$.  Note that the signs associated with $\Delta$ in the matrix above are different from the conventional definition of Gorkov's equation, because of the $\tau^3$ factor in the definition of the Green's functions.} 
\begin{equation}
\Delta = \lambda \;\underset{2 \rightarrow 1+}{\text{limit}} \; <T\hat{\psi}_{1\uparrow} \hat{\psi}_{2\downarrow}> =i \lambda  \;\underset{2 \rightarrow 1+}{\text{limit}} \; [\tilde{G}_{12}^{\alpha \alpha}]_{12}
\label{eqn4.12}
\end{equation}
in terms of the upper left component of the Green's function $\tilde{G}_{12}^{\alpha \alpha}$, which is frequently called the anomalous Green's function (or pair amplitude), and denoted by $F$.  Here $\lambda$ is the coupling constant.  For a uniform bulk superconductor $\Delta$ is real.  For a normal metal, $\lambda$ vanishes, and hence, although the pair amplitude in a normal metal may be finite, the pair potential vanishes.
Eqn.(\ref{eqn4.12}) defines a self-consistent equation for the pair potential $\Delta$; it is defined in terms of the anomalous Green's function $F$, which in turn is determined by an equation that depends on $\Delta$.  Note that unlike $[\tilde{G}_{12}^{\alpha \alpha}]_{11}$, $[\tilde{G}_{12}^{\alpha \alpha}]_{12}$ is continuous at $t_2=t_1$, so that $\Delta$ can  also be written in terms of $[\tilde{G}_{12}^{\alpha \beta}]_{12}$ or $[\tilde{G}_{12}^{\beta \alpha}]_{12}$ at $t_2=t_1$.

Still ignoring impurity scattering, the equation of motion Eqn.(\ref{eqn4.10}) can then be written in compact form as
\begin{equation}
 \left[\left(i \hat{\tau}^3 \frac{\partial}{\partial t_1} + \frac{\nabla^2_1}{2m} + \mu  +\hat{\Delta}\right) \ominus \hat{G} \right]=0 .
\label{eqn4.13}
\end{equation}
Here $\hat{\Delta}$ represents a 4x4 matrix
\begin{equation}
\hat{\Delta}=
\begin{pmatrix}
\tilde{\Delta} & 0 \\
0 & \tilde{\Delta}
\end{pmatrix},
\label{eqn4.14}
\end{equation}
where
\begin{equation}
\tilde{\Delta} =
\begin{pmatrix}
0 & \Delta \\
-\Delta^* & 0
\end{pmatrix},
\label{eqn4.15}
\end{equation}
with $\Delta$ defined by Eqn.(\ref{eqn4.12}), and
\begin{equation}
\hat{\tau}^3=
\begin{pmatrix}
\tau^3 & 0 \\
0 & \tau^3 
\end{pmatrix}.
\label{eqn4.16}
\end{equation}

Before we move on to making the quasiclassical approximation, it is useful to obtain expressions for measurable quantities in terms of the Green's functions defined in this section.  Following Eqn.(\ref{eqn2.32}), one can define a distribution function by averaging the Keldysh component of the Green's function defined in Eqn.(\ref{eqn4.9})
\begin{equation}
\tilde{h}(\vec{R},T,\vec{p}) = \frac{i}{2 \pi} \int dE \; \tilde{G}^K(\vec{R},T; \vec{p},E).
\label{eqn4.17}
\end{equation}
However, since $\tilde{G}^K$ is a 2x2 matrix, $\tilde{h}$ is also a 2x2 matrix, so that its interpretation as a simple distribution function (in the flavor of Eqn.(\ref{eqn2.32}) for the equilibrium case, for example) is not immediately clear.  A more physical interpretation can be obtained by diagonalizing the Gorkov equations given in Eqn.(\ref{eqn4.11}).\cite{smith}  The matrix on the left hand side of this equation can be diagonalized by a unitary transformation; this transformation, of course, is just the Bogoliubov-Valatin transformation, which results in a diagonal `energy' matrix with eigenvalues
\begin{equation}
E_{\vec{p}}^2=\epsilon_{\vec{p}}^2 + \Delta^2.
\label{eqn4.18}
\end{equation}
The same transformation also diagonalizes the \textit{equilibrium} distribution matrix $\tilde{h}_0$, which now has the form
\begin{equation}
\tilde{h}_0 \simeq 
\begin{pmatrix}
1-2 f_{0\uparrow}(E_{\vec{p}}) & 0 \\
0 & 2 f_{0\downarrow}(E_{\vec{p}})-1
\end{pmatrix}.
\label{eqn4.19}
\end{equation}
The top-left component is for electron-like excitations, and the bottom-right component for hole-like excitations.  This suggests that for our combined Nambu-Gorkov-Keldysh Green's functions, the equations for the electric current and thermal current should be modified to
\begin{equation}
\vec{j}(\vec{R},T)=-\frac{ie}{2}\int \frac{dE}{2 \pi} \int \frac{d^3p}{(2\pi)^3}\vec{v}_F \text{Tr} \left[\tau^3 \tilde{G}^K(\vec{R},T, \vec{p}, E)\right],
\label{eqn4.20}
\end{equation}
and
\begin{equation}
\vec{j}_{th}(\vec{R},T)=-\frac{i}{2}\int \frac{dE}{2 \pi} \int \frac{d^3p}{(2\pi)^3} E \vec{v}_F \text{Tr} \left[\tilde{G}^K(\vec{R},T, \vec{p}, E)\right].
\label{eqn4.21}
\end{equation}
Taking the trace of the matrix takes the contributions of both electrons and holes.  Note that the expression for electrical current includes a factor of $\tau^3$ in the argument of the trace, while the thermal current does not.

\section{Quasiclassical Superconducting Green's Functions}

In principle, the properties of the superconducting system may be calculated starting from Gorkov's equations.  In practice, however, such calculations are difficult in all but the simplest cases.  The problem is that Gorkov's equations contain information at length scales much finer than those of interest.  The way around this is to make the quasiclassical approximation as we did for the case of the normal Green's functions, which we proceed to do below.

Taking into account impurity scattering, the equation of motion for the superconducting Green's function can now be written as
\begin{equation}
 \left[\left(i \hat{\tau}^3 \frac{\partial}{\partial t_1} + \frac{\partial^2_1}{2m} + \mu +\hat{\Delta} - \hat{\Sigma}_{imp}\right) \ominus \hat{G} \right]=0 ,
\label{eqn5.1}
\end{equation}
where the impurity self-energy $\hat{\Sigma}_{imp}$ includes contributions from both spin-flip and spin-independent elastic scattering.

The elastic contribution to the self-energy can be written as
\begin{equation}
\hat{\Sigma}_0(\vec{p}) = N_i \int \frac{d^3p'}{(2 \pi)^3} \left| v(\vec{p}-\vec{p'})\right|^2 \hat{G}(\vec{p'}).
\label{eqn5.2}
\end{equation}
Here $v$ is the impurity potential, and $N_i$ the number of impurities per unit volume.  We assume that $v(\vec{p})$ is independent of the magnitude of $\vec{p}$, so that 
\begin{equation}
\hat{\Sigma}_0(p)=N_i N_0 \int d\epsilon_{\vec{p}'}  \; \frac{d\Omega_{p'}}{4 \pi}  \left| v(\hat{p}\cdot\hat{p}')\right|^2 \hat{G}(\vec{p}').
\label{eqn5.3}
\end{equation}
Defining the elastic scattering rate $1/\tau$ in the Born approximation by
\begin{equation}
\frac{1}{\tau}= 2 \pi N_i N_0  \int \frac{d\Omega_{p'}}{4 \pi}  \left| v(\hat{p}\cdot\hat{p}')\right|^2,
\label{eqn5.4}
\end{equation}
we can write
\begin{equation}
\hat{\Sigma}_0=\frac{1}{2 \pi \tau} \int d\epsilon_{\vec{p}} \; \hat{G}(p)
\label{eqn5.5}
\end{equation}
Here $\hat{G}(p)$ is a function of the magnitude of $\vec{p}$ alone.  Similarly, for the contribution from spin-flip scattering, one obtains
\begin{equation}
\hat{\Sigma}_{sf} = \frac{1}{2 \pi \tau_{sf}} \int d\epsilon_{\vec{p}} \; \hat{\tau}^3 \hat{G}(p) \hat{\tau}^3.
\label{eqn5.6}
\end{equation}
From Eqn.(\ref{eqn4.9}), one has
\begin{equation}
 \left[\left(i \hat{\tau}^3 \frac{\partial}{\partial t_1} + \frac{\partial^2_1}{2m} + \mu +\hat{\Delta} - \hat{\Sigma}_0 - \hat{\Sigma}_{sf}\right) \ominus \hat{G} \right]=0.
\label{eqn5.7}
\end{equation}
Converting the left hand side of this equation to relative coordinates using Eqn.(\ref{eqn2.46}) (neglecting any derivatives with respect to the center-of-mass time $T$), and using only the first term of Eqn.(\ref{eqn2.56}) to lowest order, we obtain
\begin{equation}
[\hat{\tau}^3E + \hat{\Delta}, \hat{G}] + i \vec{v}_F \cdot \partial_{\vec{R}}\hat{G} - [\hat{\Sigma}_0 + \hat{\Sigma}_{sf},\hat{G}]=0.
\label{eqn5.8}
\end{equation}
Note that there are no integrations over space or time coordinates in the equation above, but only matrix multiplications.  Written in terms of the quasiclassical Green's functions, this becomes
\begin{equation}
[\hat{\tau}^3E + \hat{\Delta}, \hat{g}] + i \vec{v}_F \cdot \partial_{\vec{R}}\hat{g} - [\hat{\sigma}_0 + \hat{\sigma}_{sf},\hat{g}]=0,
\label{eqn5.9}
\end{equation}
where
\begin{equation}
\hat{\sigma}_0=-i \frac{1}{2 \tau} \hat{g}_s, 
\label{eqn5.10}
\end{equation}
and
\begin{equation}
\hat{\sigma}_{sf} = -i\frac{1}{2 \tau_{sf}} \hat{\tau}^3 \hat{g}_s \hat{\tau}^3,
\label{eqn5.11}
\end{equation}
where the subscript to the quasiclassical Green's functions denotes that they are averaged over all angles of $\vec{p}$.  Eqn.(\ref{eqn5.9}) is Eilenberger's equation,\cite{eilenberger} and is the starting point for many calculations on superconducting systems.

To complete the transformation to quasiclassical Green's functions, we express the equations for physically observable quantities in terms of the quasiclassical Green's functions.  From Eqn.(\ref{eqn4.12}), the gap parameter $\Delta$ can be expressed as
\begin{equation}
\Delta=i \lambda [\tilde{G}_{11+}^{\alpha \beta}]_{12} = \frac{i}{2}\lambda [\tilde{G}_{11+}^K]_{12}
\label{eqn5.12}
\end{equation}
using the definition Eqn.({\ref{eqn2.18}) of the Keldysh Green's function.  In the mixed representation, this can be written as
\begin{align}
\Delta &=\frac{i\lambda}{2} \int \frac{dE}{2 \pi} \int \frac{d^3p}{(2 \pi)^3} [G^K(\vec{R}, T, \vec{p}, E)]_{12} \nonumber \\
&=N_0 \frac{\lambda}{4} \int dE \int \frac{d\Omega_p}{4 \pi} [g^K(\vec{R}, t, \hat{p}, E)]_{12} \nonumber \\
&=N_0 \frac{\lambda}{8} \int dE \int \frac{d\Omega_p}{4 \pi} \text{Tr}\big{\{} (\tau^1 - i \tau^2)g^K(\vec{R}, t, \hat{p}, E) \big{\}}  
\label{eqn5.13}
\end{align}
From Eqns.(\ref{eqn4.20}) and (\ref{eqn4.21}), the electrical current and thermal currents are given by
\begin{equation}
\vec{j}(\vec{R},T)=-\frac{eN_0}{4}\int dE \int \frac{d\Omega_p}{4\pi}\vec{v}_F \text{Tr} \left[\tau^3 \tilde{g}^K(\vec{R},T, \vec{p}, E)\right],
\label{eqn5.14}
\end{equation}
and
\begin{equation}
\vec{j}_{th}(\vec{R},T)=-\frac{N_0}{4}\int dE \int \frac{d\Omega_p}{4\pi} E \vec{v}_F \text{Tr} \left[ \tilde{g}^K(\vec{R},T, \vec{p}, E)\right].
\label{eqn5.15}
\end{equation}
One can also define an equation that gives the amount of charge associated with quasiparticle excitations.  In a superconductor in equilibrium, the number of particles and holes are equal, so the quasiparticle charge should vanish.  If there is an imbalance in the population of electrons and holes, one can obtain a \textit{charge imbalance},\cite{schmid,tinkham} usually denoted by $Q^*$, which is given by
\begin{align}
Q^* &= -\frac{ie}{2} \int \frac{dE}{2 \pi} \int \frac{d^3p}{(2\pi)^3} \text{Tr}\left[\tilde{G}^K(\vec{R},T, \vec{p}, E)\right] \nonumber \\
&=-\frac{eN_0}{4} \int dE \int \frac{d\Omega_p}{4 \pi} \text{Tr}\left[\tilde{g}^K(\vec{R},T, \vec{p}, E)\right]
\label{eqn5.16}
\end{align}
$Q^*$ is essentially the first term on the right hand side of Eqn.(\ref{eqn3.24}).  Invoking charge neutrality, this then would result in a electrochemical potential given
Eqn.(\ref{eqn3.25})
\begin{equation}
\phi(\vec{R}, T)=-\frac{1}{N_0 e} Q^*
\label{eqn5.17}
\end{equation} \\

\noindent \textit{The normalization condition and the distribution function}\\
In obtaining the equation of motion of the Green's functions by subtracting the Dyson equation from its conjugate, some information was lost regarding the norm of the quasiclassical Green's function $\hat{g}$.  The norm of $\hat{g}$ can be obtained by the normalization condition for the quasiclassical Green's function obtained by Eilenberger,\cite{eilenberger} and Larkin and Ovchinnikov\cite{larkin}
\begin{equation}
\hat{g}\hat{g}=\hat{\tau}^0,
\label{eqn5.18}
\end{equation}    
where (4x4) matrix multiplication is implied.  This normalization condition can be obtained by explicit calculation for a bulk system in equilibrium.  Furthermore, it can be shown that the normalization condition is consistent with Eilenberger's equation Eqn.(\ref{eqn5.9}).

Eqn.(\ref{eqn5.18}) is equivalent to the three (2x2) matrix equations
\begin{subequations} \label{eqn5.19}
\begin{align}
\tilde{g}^R \tilde{g}^R &=\tau^0, \label{eqn5.19a} \\
\tilde{g}^A \tilde{g}^A &= \tau^0, \label{eqn5.19b} \\
\intertext{and}
\tilde{g}^R\tilde{g}^K + \tilde{g}^K\tilde{g}^A &=0 \label{eqn5.19c}.
\end{align}
\end{subequations}

As in the case of the $E$-integrated Green's functions, we would like to obtain an equation of motion for a distribution function, eqiuvalent to the quantum Boltzmann equation derived earlier.  We introduce such a distribution function $\tilde{h}$ by the ansatz\cite{rammer}
\begin{equation}
\tilde{g}^K=\tilde{g}^R \tilde{h} - \tilde{h} \tilde{g}^A.
\label{eqn5.20}
\end{equation}
This form of the quasiclassical Keldysh Green's function satisfies the normalization condition Eqn.(\ref{eqn5.19c}), as can be verified by direct substitution
\begin{equation}
\tilde{g}^R(\tilde{g}^R \tilde{h} - \tilde{h} \tilde{g}^A) + (\tilde{g}^R \tilde{h} - \tilde{h} \tilde{g}^A)\tilde{g}^A =0
\label{eqn5.21}
\end{equation}
using Eqns.(\ref{eqn5.19a}) and(\ref{eqn5.19b}).  Note that the function $\tilde{h}$ is not uniquely defined; if $\tilde{h}$ is replaced by $\tilde{h} + \tilde{h}'$, where
\begin{equation}
\tilde{h}'=\tilde{g}^R\tilde{x} + \tilde{x}\tilde{g}^A
\label{eqn5.22}
\end{equation}
and $\tilde{x}$ is an arbitrary matrix function, the right hand side of Eqn.(\ref{eqn5.20}) is unchanged.\cite{serene}  This arbitrariness allows us some flexibility in choosing the distribution function $\tilde{h}$.  At low frequencies, for example, following Schmid and Sch\"on,\cite{schmid} and Larkin and Ovchinnikov,\cite{larkin2} we may choose $\tilde{h}$ to be diagonal in particle-hole space
\begin{equation}
\tilde{h}=h_L \tau^0 + h_T \tau^3.
\label{eqn5.23}
\end{equation}
The subscripts refer to the longitudinal ($h_L$) and transverse ($h_T$), a terminology introduced by Schmid and Sch\"on to refer to changes that are associated with the magnitude ($h_L$) or  phase ($h_T$) of the complex order parameter.  In equilibrium, $h_L(E)=1-2f_0(E)$ and $h_T(E)=0$.

Another possible choice is one introduced by Shelankov\cite{shelankov}
\begin{equation}
\tilde{h}=h_1\tau^0 + \tilde{g}^R h_2.
\label{eqn5.24}
\end{equation}
 This representation has the advantage that the equation for the distribution function reduces to a form very similar to the Boltzmann equation when the quasiparticle approximation is valid.  In what follows, we shall use the representation of Schmid and Sch\"on.
 
 The representation of the Keldysh Green's function in terms of a distribution function allows us to obtain an equation of motion for the distribution function from the equation of motion for the Green's function.  From Eilenberger's equation (\ref{eqn5.9}), we obtain three equations of motion for the three components of the quasiclassical Green's function
 \begin{subequations} \label{eqn5.25}
 \begin{align}
  [E\tau^3 +\tilde{\Delta},\tilde{g}^R]& + i\vec{v}_F \cdot \partial_{\vec{R}}\tilde{g}^R - [\tilde{\sigma}^R,\tilde{g}^R] =0 \label{eqn5.25a} \\
  [E\tau^3 +\tilde{\Delta},\tilde{g}^A]& + i\vec{v}_F \cdot \partial_{\vec{A}}\tilde{g}^A - [\tilde{\sigma}^A,\tilde{g}^A] =0 \label{eqn5.25b} \\
  [E\tau^3 +\tilde{\Delta},\tilde{g}^K]& + i\vec{v}_F \cdot \partial_{\vec{R}}\tilde{g}^K  
   -\tilde{\sigma}^R\tilde{g}^K   -\tilde{\sigma}^K\tilde{g}^A
  +\tilde{g}^R\tilde{\sigma}^K +  \tilde{g}^K\tilde{\sigma}^A \label{eqn5.25c} =0
  \end{align}
  \end{subequations}
where $\tilde{\sigma} = \tilde{\sigma}_0 + \tilde{\sigma}_{sf}$. Substituting $\tilde{g}^K$ from Eqn.(\ref{eqn5.20}) into Eqn.(\ref{eqn5.25c}), we obtain
\begin{equation}  
  [E\tau^3 +\tilde{\Delta},\tilde{g}^R \tilde{h} - \tilde{h} \tilde{g}^A] + i\vec{v}_F \cdot \partial_{\vec{R}}(\tilde{g}^R \tilde{h} - \tilde{h} \tilde{g}^A)  
   -\tilde{\sigma}^R(\tilde{g}^R \tilde{h} - \tilde{h} \tilde{g}^A)   -\tilde{\sigma}^K\tilde{g}^A
  +\tilde{g}^R\tilde{\sigma}^K + (\tilde{g}^R \tilde{h} - \tilde{h} \tilde{g}^A)\tilde{\sigma}^A  =0.
\label{eqn5.26}
\end{equation}
Subtracting from the equation above Eqn.(\ref{eqn5.25a}) multiplied by $\tilde{h}$ on the right, and adding Eqn.(\ref{eqn5.25b}) multiplied by $\tilde{h}$ on the left, we obtain
\begin{equation}
\tilde{g}^R B[\tilde{h}] - B[\tilde{h}]\tilde{g}^A=0,
\label{eqn5.27}
\end{equation}
where 
\begin{equation}
B[\tilde{h}]= [E\tau^3 +\tilde{\Delta},\tilde{h}]+\tilde{\sigma}^K -(\tilde{\sigma}^R \tilde{h} - \tilde{h}\tilde{\sigma}^A) + i \vec{v}_F \cdot \partial_{\vec{R}}\tilde{h}.
\label{eqn5.28}
\end{equation}
Equation(\ref{eqn5.27}) is the required equation of motion for the distribution function $\tilde{h}$.

As an example, we shall calculate the Green's functions for the equilibrium case for a bulk superconductor, in the limit where $\sigma=0$.  Equation(\ref{eqn5.25a}) in this limit has the form
\begin{equation}
[E\tau^3 + \tilde{\Delta},\tilde{g}^R]=0
\label{eqn5.29}  
\end{equation}
First, if we represent the retarded Green's function by 
\begin{equation}
\tilde{g}^R =
\begin{pmatrix}
g_{11} & g_{12} \\
g_{21} & g_{22} 
\end{pmatrix},
\label{eqn5.30}
\end{equation} 
the normalization condition Eqn.(\ref{eqn5.19a}) and Eqn.(\ref{eqn5.29}) together imply that 
\begin{subequations} \label{eqn5.31}
\begin{align}
&g_{11} =-g_{22}, \label{eqn5.31a} \\
&g_{11}^2 + g_{21}g_{12}=1, \label{eqn5.31b} \\
&g_{21}=-g_{12}\frac{\Delta^*}{\Delta} \label{eqn5.31c} \\
\intertext{and}
&g_{11}=g_{12}\frac{E}{\Delta}. \label{eqn5.3d}
\end{align}
\end{subequations}
Solving these equations, we obtain
\begin{equation}
\tilde{g}^R = 
\begin{pmatrix}
\frac{E}{\sqrt{E^2- |\Delta|^2}} & \frac{\Delta}{\sqrt{E^2- |\Delta|^2}} \\
-\frac{\Delta^*}{\sqrt{E^2- |\Delta|^2}} & -\frac{E}{\sqrt{E^2- |\Delta|^2}}
\end{pmatrix}.
\label{eqn5.32}
\end{equation}
In taking the square roots in the equation above, a factor of $i$ will appear if $E^2 < |\Delta|^2$.  This is the expected solution from Gorkov's equations.  Note that $g_{11}$ is just the normalized BCS density of states for $E^2 > |\Delta|^2$, so that $g_{11}$ represents a generalized density of states.

From the equations above, it is clear that $g_{21}=-g^*_{12}$.  To obtain $\tilde{g}^A$, we can use the relation Eqn.(\ref{eqn2.7}), which in terms of Nambu-Gorkov matrices reads
\begin{equation}
\tilde{G}^A = -\tau^3 \tilde{G}^{R+}\tau^3 \qquad \text{or} \qquad \tilde{g}^A = -\tau^3 \tilde{g}^{R+}\tau^3
\label{eqn5.33}
\end{equation}
where the $^+$ symbol denotes the Hermitian conjugate.

\section{The dirty limit:  the Usadel equation}
In systems where elastic impurity scattering is strong, the motion of quasiparticles is not ballistic, but diffusive.  In this case, the effect of the strong impurity scattering is to randomize the momentum of the quasiparticles, so that it makes sense to average the properties of system over the directions of the momentum, keeping in mind that the magnitude of the momentum is still its value at the Fermi surface, $p_F$.  

With this in mind, let us consider a system with strong impurity scattering, which we define as a system in which the impurity scattering rate $1/\tau$, as defined by Eqn.(\ref{eqn5.4}), is much larger than any energy in the problem ($E, \Delta$).  In this case, we expand the quasiclassical Green's function to first order in momentum (essentially an expansion in spherical harmonics)
\begin{equation}
\hat{g}=\hat{g}_s + \hat{p}\hat{g}_p
\label{eqn6.1}
\end{equation}
where $\hat{p}$ denotes a vector of unit magnitude in the direction of $\vec{p}$, and $\hat{g}_s$ and $\hat{g}_p$ are independent of the direction of $\vec{p}$, and the subscripts stand for $s$-wave and $p$-wave components of $\hat{g}$.  The assumption is that $\hat{g}_p \ll \hat{g}_s$.  The self-energy is expanded in a similar fashion
\begin{equation}
\hat{\sigma}=\hat{\sigma}_0 + \hat{\sigma}_{sf}=\hat{\sigma}_s + \hat{p}\hat{\sigma}_p
\label{eqn6.2}
\end{equation}
We would like to calculate the components of $\hat{\sigma}$ in terms of the components of $\hat{g}$, for which we shall need the definition of $\hat{\sigma}$ in terms of $\hat{g}$.  Ignoring spin-flip scattering for the moment, this relation is given by Eqn.(\ref{eqn5.3}), which can be rewritten in terms of the quasiclassical Green's functions in the form
\begin{equation}
\hat{\sigma}_0(p)=-i \pi N_i N_0 \int \frac{d\Omega_{p'}}{4 \pi}|v(\hat{p}\cdot \hat{p}')|^2 \hat{g}(\vec{p}') = -i \pi N_i N_0 \int \frac{d\Omega_{p'}}{4 \pi}|v(\hat{p}\cdot \hat{p}')|^2 (\hat{g}_s + \hat{p}'\hat{g}_{p'}),
\label{eqn6.3}
\end{equation}
using the expansion Eqn.(\ref{eqn6.1}) for the Green's function above.
 In order to do this, we take the dot product of  $\hat{p}$ with both sides of the equation above
\begin{equation}
\hat{p}\cdot\hat{\sigma}=\hat{p}\hat{\sigma}_s + (\hat{p}\cdot\hat{p})\hat{\sigma}_p = \hat{p}\hat{\sigma}_s + \hat{\sigma}_p.
\label{eqn6.4}
\end{equation}
We perform a similar operation on Eqn.(\ref{eqn6.3})
\begin{equation}
\hat{p}\cdot \hat{\sigma}_0(p)= -i \pi N_i N_0 \int \frac{d\Omega_{p'}}{4 \pi}|v(\hat{p}\cdot \hat{p}')|^2 (\hat{p}\hat{g}_s + (\hat{p}\cdot\hat{p}')\hat{g}_{p'}).
\label{eqn6.5}  
\end{equation}
Since both $\hat{p}$ and $\hat{g}_s$ are independent of the integration over $d\Omega_{p'}$, the first term under the integral in the equation above can be written as
\begin{equation}
-i \pi \hat{p} \hat{g}_s N_i N_0 \int \frac{d\Omega_{p'}}{4 \pi} |v(\hat{p}\cdot \hat{p}')|^2 = - \frac{i}{2 \tau} \hat{p}\hat{g}_s,
\label{eqn6.6}
\end{equation}
using the definition Eqn.(\ref{eqn5.4}) of $1/\tau$.  If we consider non-spin-flip scattering alone, then $\hat{\sigma}$ can be obtained by equating like terms in Eqns.(\ref{eqn6.4}) and (\ref{eqn6.5}).  If we include spin-flip scattering, one can write
\begin{equation}
\hat{\sigma}_s = - \frac{i}{2 \tau}\hat{g}_s - - \frac{i}{2 \tau_{sf}} \hat{\tau}^3 \hat{g}_s \hat{\tau}^3
\label{eqn6.7}
\end{equation}
as expected from Eqns.(\ref{eqn5.10}) and (\ref{eqn5.11}).  For the $p$-wave component, we consider only the contribution from non spin-flip impurity scattering, under the assumption that it is much stronger than the spin-flip scattering.  The second term under the integral in Eqn.(\ref{eqn6.5}) can be written as
\begin{equation}
 -i \pi N_i N_0 \int \frac{d\Omega_{p'}}{4 \pi}|v(\hat{p}\cdot \hat{p}')|^2 (\hat{p}\cdot\hat{p}')\hat{g}_{p'} =   -i \pi N_i N_0 \hat{g}_{p}\int \frac{d\Omega_{p'}}{4 \pi}|v(\hat{p}\cdot \hat{p}')|^2 [1-(1-\hat{p}\cdot\hat{p}')],
\label{eqn6.8}
\end{equation}
where $\hat{g}_p$ can be taken out of the integral since it is independent of the direction of $\vec{p}$.   The first term in the square bracket can be seen to be related to the elastic scattering rate $1/\tau$.  The remaining terms can be written in terms of the transport time, defined by
\begin{equation}
\frac{1}{\tau_{tr}} =  2 \pi N_i N_0 \int \frac{d\Omega_{p'}}{4 \pi}|v(\hat{p}\cdot \hat{p}')|^2 (1-\hat{p}\cdot\hat{p}'),
\label{eqn6.9}
\end{equation}
that is well known in the transport theory of metals.  With this definition, $\hat{\sigma}_p$ can be written as
\begin{equation}
\hat{\sigma}_p=-\frac{i}{2}(\frac{1}{\tau} - \frac{1}{\tau_{tr}}) \hat{g}_p.
\label{eqn6.10}
\end{equation} 
By putting Eqn.(\ref{eqn6.1}) into the normalization condition for the quasiclassical Green's function, and neglecting terms quadratic in $\hat{g}_p$, we also obtain the two equations
\begin{subequations} \label{eqn6.11}
\begin{align}
&\hat{g}_s \hat{g}_s = 1 \label{eqn6.11a} \\
\intertext{and}
&\hat{g}_s \hat{g}_p + \hat{g}_p \hat{g}_s = 0 \label{eqn6.11b}.
\end{align}
\end{subequations}
We now proceed to expand the Eilenberger equation, Eqn.(\ref{eqn5.9}), in terms of the $s$ and $p$-wave expansions of $\hat{g}$ and $\hat{\sigma}$.  Replacing $\vec{v}_F$ by $v_F \hat{p}$, we obtain
\begin{equation}
[\hat{\tau}^3E -\hat{\Delta},\hat{g}_s] + \hat{p}[\hat{\tau}^3E -\hat{\Delta}, \hat{g}_p] + i v_F \hat{p} \cdot \partial_{\vec{R}}(\hat{g}_s + \hat{p}\hat{g}_p) - [\hat{\sigma}_s,\hat{g}_s] - \hat{p}{[\hat{\sigma}_s,\hat{g}_p] + [\hat{\sigma}_p,\hat{g}_s]} + \hat{p} \cdot \hat{p} [\hat{\sigma}_p,\hat{g}_p]= 0.
\label{eqn6.12}
\end{equation}
The last term is second order in the small quantity $\hat{g}_p$, and can be neglected.  Collecting the terms that are even in $\hat{p}$, we obtain
\begin{equation}
[\hat{\tau}^3E + \hat{\Delta},\hat{g}_s] + iv_F (\hat{p} \cdot \hat{p}) \partial_{\vec{R}} \hat{g}_p =0.
\label{eqn6.13}
\end{equation}
Averaging this equation over all directions of $\hat{p}$ gives
\begin{equation}
[\hat{\tau}^3E + \hat{\Delta},\hat{g}_s] + i\frac{v_F}{3} \partial_{\vec{R}} \hat{g}_p =0.
\label{eqn6.14}
\end{equation}
Ignoring spin-dependent scattering for the moment, the terms that are odd in $\hat{p}$ can be written
\begin{equation}
[\hat{\tau}^3E + \hat{\Delta},\hat{g}_p]+ i v_F \partial_{\vec{R}}\hat{g}_s - \frac{i}{2\tau_{tr}}[\hat{g}_p,\hat{g}_s] = 0,
\label{eqn6.15}
\end{equation}
where we have used Eqns.(\ref{eqn6.7}) and (\ref{eqn6.10}) to write $\hat{\sigma}_s$ and $\hat{\sigma}_p$ in terms of $\hat{g}_s$ and $\hat{g}_p$.  If elastic scattering is strong, the first term in the equation above can be neglected compared to the third, so we obtain
\begin{equation}
i v_F \partial_{\vec{R}}\hat{g}_s + \frac{i}{\tau_{tr}}\hat{g}_s\hat{g}_p = 0,
\label{eqn6.16}
\end{equation}
where we have used Eqn.(\ref{eqn6.11b}) to simplify the second term.  Multiplying this equation by $\hat{g}_s$ on the left, and using Eqn.(\ref{eqn6.11a}), we obtain
\begin{equation}
v_F\hat{g}_s \partial_{\vec{R}}\hat{g}_s = -\frac{1}{\tau_{tr}}\hat{g}_p,
\label{eqn6.17}
\end{equation}
or writing $\hat{g}_p$ in terms of $\hat{g}_s$
\begin{equation}
\hat{g}_p = v_F \tau_{tr} \hat{g}_s \partial_{\vec{R}}\hat{g}_s = -\ell \hat{g}_s \partial_{\vec{R}}\hat{g}_s
\label{eqn6.18}
\end{equation}
where we have introduced the elastic scattering length $\ell=v_F \tau_{tr}$.  Putting this into Eqn.(\ref{eqn6.15}), we obtain
\begin{equation}
[\hat{\tau}^3E +\hat{\Delta},\hat{g}_s] - i\frac{v_F \ell}{3} \partial_{\vec{R}} \hat{g}_s \partial_{\vec{R}}\hat{g}_s =0,
\label{eqn6.19}
\end{equation}
or writing this in terms of the diffusion coefficient $D=(1/3) v_F \ell$, and reintroducing the spin-flip scattering term,
\begin{equation}
[\hat{\tau}^3E + \hat{\Delta}-\hat{\sigma}_{sf},\hat{g}_s] - iD \partial_{\vec{R}} \hat{g}_s \partial_{\vec{R}}\hat{g}_s =0.
\label{eqn6.20}
\end{equation}
This is the equation first derived by Usadel,\cite{usadel} and forms the starting point for most discussions on dirty superconducting systems.

In the remainder of our development, we shall neglect the spin-flip scattering term.  Writing $\hat{g}_s$ as a matrix
\begin{equation}
\hat{g}_s = 
\begin{pmatrix}
\tilde{g}_s^R & \tilde{g}_s^K \\
0 & \tilde{g}_s^A
\end{pmatrix},
\label{eqn6.21}
\end{equation}
we can write the matrix equation, Eqn.(\ref{eqn6.20}), as three separate equations
\begin{subequations} \label{eqn6.22}
\begin{align}
[\tau^3E +\tilde{\Delta},\tilde{g}_s^R] &= iD\partial_{\vec{R}}(\tilde{g}_s^R \partial_{\vec{R}}\tilde{g}_s^R), \label{eqn6.22a} \\ 
[\tau^3E + \tilde{\Delta},\tilde{g}_s^A] &= iD\partial_{\vec{R}}(\tilde{g}_s^A \partial_{\vec{R}}\tilde{g}_s^A)
\label{eqn6.22b}, \\
\intertext{and}
[\tau^3E + \tilde{\Delta},\tilde{g}_s^K] &= iD\partial_{\vec{R}}\left[(\tilde{g}_s^R \partial_{\vec{R}}\tilde{g}_s^K)+
(\tilde{g}_s^K \partial_{\vec{R}}\tilde{g}_s^A)\right].
\label{eqn6.22c}
\end{align}
\end{subequations}
As before, the first two equations come from the diagonal components of the Usadel equation, and describe the equilibrium properties of the system, while the third equation comes from the off-diagonal or Keldysh component, and represents the kinetic equation for the distribution function.  These equations are supplemented by the equations for measurable quantities corresponding to Eqns.(\ref{eqn5.14}), (\ref{eqn5.15}) and (\ref{eqn5.14}).  Expanding $\tilde{g}^K$ in Eqn.(\ref{eqn5.14} using Eqns.(\ref{eqn6.1}) and Eqns.(\ref{eqn5.14}), we have
\begin{equation}
\vec{j}(\vec{R},T)=-\frac{eN_0}{4}\int dE \int \frac{d\Omega_p}{4\pi}v_F \hat{p} \text{Tr} \left[\tau^3 \left(\tilde{g}_s^K + \hat{p}\tilde{g}_p^K\right) \right]=\frac{eN_0D}{4}\int dE \text{Tr} \left[\tau^3 (\tilde{g}_s \partial_{\vec{R}}\tilde{g}_s)^K \right],
\label{eqn6.23}
\end{equation}
where the angular average over the $p$-wave component give a factor of $(1/3)$, and we have replaced $(1/3) v_F \ell$ by the diffusion coefficient $D$.  Now 
\begin{equation}
(\tilde{g}_s \partial_{\vec{R}}\tilde{g}_s)^K = (\tilde{g}_s^R \partial_{\vec{R}}\tilde{g}_s^K)+
(\tilde{g}_s^K \partial_{\vec{R}}\tilde{g}_s^A)
\label{eqn6.24}
\end{equation}
so that Eqn.(\ref{eqn6.24}) above can be rewritten as
\begin{equation}
\vec{j}(\vec{R},T)=\frac{eN_0D}{4}\int dE \text{Tr} \left[\tau^3 \left(\tilde{g}_s^R \partial_{\vec{R}}\tilde{g}_s^K+
\tilde{g}_s^K \partial_{\vec{R}}\tilde{g}_s^A \right) \right].
\label{eqn6.25}
\end{equation} 
The equation for $j_{th}$ is obtained in a similar way
\begin{equation}
\vec{j}_{th}(\vec{R},T)=\frac{N_0D}{4}\int dE \;E \;\text{Tr} \left[ \tilde{g}_s^R \partial_{\vec{R}}\tilde{g}_s^K +
\tilde{g}_s^K \partial_{\vec{R}}\tilde{g}_s^A \right],
\label{eqn6.26}
\end{equation}
   
The kinetic equation, Eqn.(\ref{eqn6.22c}), can be recast in terms of a differential equation for a distribution function $\tilde{h}$, in the hopes of separating the equilibrium properties of the system, represented by $\tilde{g}_s^R$ and $\tilde{g}_s^A$, from the non-equilibrium properties of the system, represented by $\tilde{h}$.  To this end, as before, we substitute Eqn.(\ref{eqn5.20}) for $\tilde{g}_s^K$ in Eqn.(\ref{eqn6.22c}), which then becomes
\begin{equation}
[\tau^3E +\tilde{\Delta}, \tilde{g}_s^R\tilde{h}] - [\tau^3E +\tilde{\Delta}, \tilde{h}\tilde{g}_s^A]= iD\partial_{\vec{R}}\left[\left(\tilde{g}_s^R(\partial_{\vec{R}}\tilde{g}_s^R)\tilde{h} - \tilde{h}\tilde{g}_s^A(\partial_{\vec{R}}\tilde{g}_s^A)\right) + \partial_{\vec{R}}\tilde{h} - \left(\tilde{g}_s^R\partial_{\vec{R}}\tilde{h}\tilde{g}_s^A\right) \right].
\label{eqn6.27}
\end{equation}
Taking the trace of both sides of the equation above gives
\begin{equation}
0=iD\partial_{\vec{R}}\text{Tr}\Big{\{}\partial_{\vec{R}}\tilde{h} +
\tilde{g}_s^R(\partial_{\vec{R}}\tilde{g}_s^R)\tilde{h} -
\tilde{h}\tilde{g}_s^A(\partial_{\vec{R}}\tilde{g}_s^A) -
\tilde{g}_s^R(\partial_{\vec{R}}\tilde{h}) \tilde{g}_s^A \Big{\}},
\label{eqn6.28}
\end{equation}
using Eqns.(\ref{eqn6.22a}) and (\ref{eqn6.22b}).  In the linear regime, we can use the diagonal representation of $\tilde{h}$ given by Eqn.(\ref{eqn5.23}) to obtain
\begin{equation}
D \partial_{\vec{R}} \left[ (\partial_{\vec{R}}h_L) \text{Tr}\Big{\{}1-\tilde{g}_s^R\tilde{g}_s^A\Big{\}} + h_T \text{Tr}\Big{\{}\tau^3\left(\tilde{g}_s^R(\partial_{\vec{R}}\tilde{g}_s^R)-\tilde{g}_s^A(\partial_{\vec{R}}\tilde{g}_s^A) \right)\Big{\}} - (\partial_{\vec{R}}h_T)\text{Tr}\Big{\{}\tilde{g}_s^R \tau^3 \tilde{g}_s^A\Big{\}}\right]=0,
\label{eqn6.29}
\end{equation}
where we have used the fact that
\begin{equation}
\text{Tr}\big{\{}\partial_{\vec{R}}(\tilde{g}_s^R\tilde{g}_s^R)\big{\}}=
\text{Tr}\big{\{}(\partial_{\vec{R}}(\tilde{g}_s^R)\tilde{g}_s^R + 
\tilde{g}_s^R(\partial_{\vec{R}}(\tilde{g}_s^R)\big{\}}= 2\text{Tr}\big{\{}\tilde{g}_s^R(\partial_{\vec{R}}(\tilde{g}_s^R)\big{\}}= 0,
\label{eqn6.30}
\end{equation}
Taking the trace after mulitiplying both sides of Eqn.(\ref{eqn6.27}) by  $\tau^3$ gives 
\begin{equation}
\begin{split}
D \partial_{\vec{R}} \left[
\partial_{\vec{R}} h_T \text{Tr}
 \big{\{} 
 1-\tilde{g}_s^R 
 \tau^3 
 \tilde{g}_s^A 
 \tau^3 
 \big{\}} 
+h_L \text{Tr} \big{\{}\tau^3
\left(
\tilde{g}_s^R(\partial_{\vec{R}}\tilde{g}_s^R)-\tilde{g}_s^A(\partial_{\vec{R}}\tilde{g}_s^A)\right)\big{\}} - 
\partial_{\vec{R}}h_L \text{Tr}
\big{\{}\tilde{g}_s^R\tilde{g}_s^A \tau^3 \big{\}} 
\right]  \\
=i \left[
h_L \text{Tr}
\big{\{}\tau^3[\tilde{g}_s^R-\tilde{g}_s^A, \tilde{\Delta}] \big{\}} 
-2 h_T \text{Tr}
\big{\{}\tilde{\Delta}(\tilde{g}_s^R+\tilde{g}_s^A)\big{\}} 
\right].
\end{split}
\label{eqn6.31}
\end{equation}
Equations (\ref{eqn6.29}) and (\ref{eqn6.31}) form a set of coupled differential equations for the distribution functions $h_L$ and $h_T$.  Let us define the quantities
\begin{align}
Q&=\frac{1}{4} \text{Tr} \big{\{}  \tau^3 \left(\tilde{g}_s^R(\partial_{\vec{R}}\tilde{g}_s^R)-\tilde{g}_s^A(\partial_{\vec{R}}\tilde{g}_s^A) \right)  \big{\}} \label{eqn6.32}
\intertext{and}
M_{ij}&=\frac{1}{4}  \text{Tr} \big{\{}  \delta_{ij}\tau^0 - \tilde{g}_s^R \tau^i \tilde{g}_s^A \tau^j  \big{\}}.  \label{eqn6.33}
\end{align}
Then Eqns.(\ref{eqn6.29}) and (\ref{eqn6.31}) can be written in the form
\begin{subequations} \label{eqn6.34}
\begin{align}
\partial_{\vec{R}} \left[M_{00}(\partial_{\vec{R}}h_L) + Q h_T + M_{30}\partial_{\vec{R}}h_T) \right]&=0, \label{eqn6.34a} \\
\partial_{\vec{R}} \left[M_{33}(\partial_{\vec{R}}h_T) + Q h_L + M_{03}\partial_{\vec{R}}h_L) \right]&=\frac{i}{4D}  \left[
h_L \text{Tr}
\big{\{}\tau^3[\tilde{g}_s^R-\tilde{g}_s^A, \tilde{\Delta}] \big{\}} 
-2 h_T \text{Tr}
\big{\{}\tilde{\Delta}(\tilde{g}_s^R+\tilde{g}_s^A)\big{\}} 
\right].
\label{eqn6.34b}
\end{align}
\end{subequations}
These equations are in the form of diffusion equations for the distribution function, more general forms of the diffusion equation discussed for the normal case in the introduction.  As we shall see, the quantity $Q$ is related to the \textit{spectral} supercurrent in the system, $DM_{ij}$ are now the energy and position dependent diffusion coefficients, and Eqns.(\ref{eqn6.34}) are essentially continuity equations for the spectral thermal and electric current.  In the normal limit, $g^R=\tau^3$, $g^A=-\tau^3$, and $\tilde{\Delta}=0$, so that $M_{00}=M_{33}=1$, and $Q=M_{03}=M_{30}=0$.  Equations (\ref{eqn6.34}) then reduce to Eqn.(\ref{eqn1.1}), as expected.  Noting that the term in square brackets in Eqn.(\ref{eqn6.22c}) is the same as the term in parenthesis in Eqn.(\ref{eqn6.25}) and the term in square brackets in Eqn.(\ref{eqn6.26}), the electric current can be written as
\begin{equation}
\vec{j}(\vec{R},T)=eN_0D \int dE \; (M_{33}(\partial_{\vec{R}}h_T) + Q h_L + M_{03}\partial_{\vec{R}}h_L).
\label{eqn6.35}
\end{equation}
The first term corresponds to quasiparticle (or dissipative) current, and the second term to the supercurrent.  The third term, which is proportional to the derivative of $h_L$, is associated with an imbalance between particles and holes.  The thermal current can be written in a similar way
\begin{equation}
\vec{j}_{th}(\vec{R},T)=N_0D \int dE \;E [M_{00}(\partial_{\vec{R}}h_L) + Q h_T + M_{30}\partial_{\vec{R}}h_T].
\label{eqn6.36}
\end{equation}
For the charge-imbalance $Q^*$, we note that only the $s$-wave part of the Keldysh Green's function in the square brackets in Eqn.(\ref{eqn5.16}) survives after angular averaging.  Writing $g_s^K$ in the form given by Eqn.(\ref{eqn5.20}), with $\tilde{h}$ given by Eqn.(\ref{eqn5.23}), we have
\begin{equation}
Q^*=-\frac{eN_0}{4} \int dE  \; h_L \text{Tr}\big{\{}g_s^R-g_s^A\big{\}} - \frac{eN_0}{4} \int dE \;  h_T \text{Tr}\big{\{}\tau^3(g_s^R-g_s^A)\big{\}} = -eN_0 \int dE  \; h_T N(E),
\label{eqn6.37}
\end{equation}
since $g_s^R$ and $g_s^A$ are traceless.  Here we have defined the normalized superconducting density of states by 
\begin{equation}
N(E)=\frac{1}{4}\text{Tr} \big{\{} \tau^3 (g_s^R-g_s^A) \big{\}},
\label{eqn6.38}
\end{equation}
which reduces to the conventional BCS density of states in the equilibrium case.  

We can also recast the kinetic equations Eqn.(\ref{eqn6.34}) in a slightly different form sometimes used in the literature.  To do this, we subtract Eqn.(\ref{eqn6.22b}) from Eqn.(\ref{eqn6.22a}), multiply by $\tau^3$, and take the trace.  The result is
\begin{equation}
\text{Tr} \big{\{}\tau^3[g_s^R-g_s^A,\tilde{\Delta}] \big{\}} = - iD \partial_{\vec{R}}(\text{Tr} \big{\{} \tau^3 [\tilde{g}_s^R \partial_{\vec{R}}\tilde{g}_s^R-\tilde{g}_s^A \partial_{\vec{R}}\tilde{g}_s^A]\big{\}}).
\label{eqn6.39}
\end{equation}
Using the definition of $Q$, we have
\begin{equation}
\partial_{\vec{R}}Q=\frac{i}{4D}\text{Tr} \big{\{}\tau^3[g_s^R-g_s^A,\tilde{\Delta}] \big{\}}.
\label{eqn6.40}
\end{equation}
If we choose a gauge in which $\Delta$ is real, the right hand side of the equation vanishes, so that $\partial_{\vec{R}}Q=0$.  Clearly, $\partial_{\vec{R}}Q=0$ also for a normal metal (even a proximity-coupled normal metal), where $\Delta$ vanishes.  In either case, the spectral supercurrent $Q$ is conserved.  
The right hand side of the equation above multiplied by $h_L$  is the same as the second term of Eqn.(\ref{eqn6.34b}).  Subtracting Eqn.(\ref{eqn6.39}) multiplied by $h_L$ from Eqn.(\ref{eqn6.34b}), we obtain
\begin{equation}
\partial_{\vec{R}} \left[M_{33}(\partial_{\vec{R}}h_T)+ M_{03}(\partial_{\vec{R}}h_L) \right] + Q(\partial_{\vec{R}}h_L)=-\frac{i}{4D}  \left[
2 h_T \text{Tr}
\big{\{}\tilde{\Delta}(\tilde{g}_s^R+\tilde{g}_s^A)\big{\}} 
\right].
\label{eqn6.41}
\end{equation}
For real $\Delta$, when $\partial_{\vec{R}}Q=0$, Eqn.(\ref{eqn6.34a}) can be written in a similar manner
\begin{equation}
\partial_{\vec{R}} \left[M_{00}(\partial_{\vec{R}}h_L)+ M_{30}(\partial_{\vec{R}}h_T) \right] + Q(\partial_{\vec{R}}h_T)=0
\label{eqn6.42}
\end{equation}

Finally, we can also write the kinetic equations in a form similar to Eqn.(\ref{eqn5.27}) 
\begin{equation}
\tilde{g}_s^R B[\tilde{h}] - B[\tilde{h}] \tilde{g}_s^A =0,
\label{eqn6.43}
\end{equation}
by performing similar manipulations on Eqns.(\ref{eqn6.22}) as were performed on Eqns.(\ref{eqn5.25}).  
For the diffusive case, $B[\tilde{h}]$ is a functional of $\tilde{h}$ now given by
\begin{equation}
B[\tilde{h}]=[\tau^3E +  \tilde{\Delta}, \tilde{h}] - iD\left[(\partial_{\vec{R}}\tilde{g}_s^R) (\partial_{\vec{R}}\tilde{h}) + \frac{1}{2} \left(\tilde{g}_s^R (\partial_{\vec{R}}^2\tilde{h}) -(\partial_{\vec{R}}^2\tilde{h})\tilde{g}_s^A \right) - (\partial_{\vec{R}}\tilde{h}) (\partial_{\vec{R}}\tilde{g}_s^A)\right].
\label{eqn6.44}
\end{equation} \\

\noindent \textit{Boundary conditions for the quasiclassical equations of motion} \\
Eqns.(\ref{eqn6.22}) or their derivatives form a set of coupled differential equations for the quasiclassical Green's functions and the distribution function.  In order to obtain a solution, however, we need to specify the boundary conditions for the Green's functions and the distribution function.  To this end, we define the concept of a \textit{reservoir}, where the Green's functions and distribution function have well-defined values.  

For a normal reservoir, the retarded and advanced quasiclassical Green's functions are given by 
\begin{equation}
g_{N0}^R=\tau^3;  \qquad   g_{N0}^A=-\tau^3,
\label{eqn6.45}
\end{equation}
and for the superconducting case, by Eqn.(\ref{eqn5.32}), which we reproduce here
\begin{equation}
\tilde{g}^R = 
\begin{pmatrix}
\frac{E}{\sqrt{E^2- |\Delta|^2}} & \frac{\Delta}{\sqrt{E^2- |\Delta|^2}} \\
-\frac{\Delta^*}{\sqrt{E^2- |\Delta|^2}} & -\frac{E}{\sqrt{E^2- |\Delta|^2}}
\end{pmatrix}.
\label{eqn6.46}
\end{equation}
with $g_{S0}^A$ given by Eqn.(\ref{eqn5.33}).

The equilibrium distribution function $\tilde{h}$ is given by Eqn.(\ref{eqn4.19}) in both normal and superconducting reservoirs (since we are dealing only with excitations), where, as we noted earlier the $(1,1)$ component of the matrix applies to particle-like excitations, and the $(2,2)$ component of the matrix applies to hole-like excitations.  From this point of view, the Fermi functions in Eqn.(\ref{eqn4.19}) are given in terms of the usual equilibrium Fermi function Eqn.(\ref{eqn1.5}) by $f_{0\uparrow}(E)=f_0(E)$ and $f_{0\downarrow}(E)=f_0(-E)$.\footnote{Since we are dealing with the static limit in all that follows, we again use the symbol $T$ to refer to the temperature}  Looking ahead to where we might have a finite voltage $V$ on a reservoir, the equilibrium form of $\tilde{h}$ can then be written as
\begin{equation}
\tilde{h}_0=
\begin{pmatrix}
\tanh\left(\frac{E+eV}{2 k_B T}\right) & 0 \\
0 & \tanh\left(\frac{E-eV}{2 k_B T}\right)
\end{pmatrix}
\label{eqn6.47}
\end{equation}
If we write $\tilde{h}_0$ in the form of Eqn.(\ref{eqn5.23}) the equilibrium values of $h_L$ and $h_T$ can then be expressed as
\begin{equation}
h_{L,T}=\frac{1}{2}\left[\tanh\left(\frac{E+eV}{2 k_B T}\right) \pm \tanh\left(\frac{E-eV}{2 k_B T}\right)\right]
\label{eqn6.48}
\end{equation}
If a finite voltage is put on the superconductor, we will obtain a time evolution of the phase in accordance with the Josephson relations.  Since we have restricted ourselves here to the static case, we must assume that the voltage on the superconducting reservoir $V=0$.  In this case, $h_T=0$ for the superconducting reservoir.

For a system with perfect interfaces between the superconducting and normal parts, the boundary conditions 
defined by the equations above are sufficient to solve the differential equations, the implicit assumption being that the Green's functions and the distribution functions are continuous across any interface.  When the transparency of the interface is less than unity, this is no longer true.  Zaitsev\cite{zaitsev} derived the boundary conditions for the Green's functions at an interface of arbitrary transparency. Kupriyanov and Lukichev\cite{kuprianov} 
simplified these equations for the diffusive case in the limit of small barrier transparency.  Consider then an interface at $x=0$ between two metals, say one a normal metal in the half-plane $x<0$ (labeled by the index `1') and one a superconductor in the half-plane $x>0$, (labeled by the index `2'), although it also could be an interface between two different normal metals or superconductors.  The boundary conditions of Kupriyanov and Lukichev are then
\begin{subequations} \label{eqn6.49}
\begin{align}
v_{F1} D_1 \hat{g}_{s1} (\partial_x \hat{g}_{s1}) &= v_{F2} D_2 \hat{g}_{s2} (\partial_x \hat{g}_{s2}), \label{eqn6.49a} \\
\hat{g}_{s1}\partial_x\hat{g}_{s1} &= \frac{1}{2r}[\hat{g}_{s1},\hat{g}_{s2}] \label{eqn6.49b}.
\end{align}
\end{subequations}
Here $\partial_x$ denotes a derivative in the positive $x$ direction, and $r=R_b/R_N$ is a parameter that is nominally the ratio of the barrier resistance $R_b$ to the normal wire resistance per unit length $R_N/L$, but is inversely proportional to the transmission $t$ of the interface.  The first of the equations is clearly related to the conservation of current across the interface.  The right hand side of the second equation has been shown to be the first term in an expansion of terms of the transmission coefficient $t$,\cite{lambert} and hence, it is only valid for low $t$.  

The diagonal part of Eqn.(\ref{eqn6.49b}) gives the boundary condition for the Green's functions
\begin{equation}
\tilde{g}_{s1}^R \partial_x \tilde{g}_{s1}^R = \frac{1}{2r}\left[\tilde{g}_{s1}^R \tilde{g}_{s2}^R - \tilde{g}_{s2}^R \tilde{g}_{s1}^R \right]
\label{eqn6.50}
\end{equation}
The off-diagonal part of Eqn.(\ref{eqn6.49b}) gives the boundary condition for the distribution function
\begin{equation}
\tilde{g}_{s1}^R \partial_x \tilde{g}_{s1}^K + \tilde{g}_{s1}^K \partial_x \tilde{g}_{s1}^A =
 \frac{1}{2r}\left[\tilde{g}_{s1}^R \tilde{g}_{s2}^K +\tilde{g}_{s1}^K \tilde{g}_{s2}^A- \tilde{g}_{s2}^R \tilde{g}_{s1}^K - \tilde{g}_{s2}^K \tilde{g}_{s1}^A  \right]
\label{eqn6.51}
\end{equation}
If we put in Eqn.(\ref{eqn5.20}) for $\tilde{g}_s^K$, with Eqn.(\ref{eqn5.23}) for $\tilde{h}$, and then take as before the trace of the resulting equation, and the trace of the equation multiplied by $\tau^3$, we will obtain boundary conditions for $h_T$ and $h_L$.  Noting that the left hand side of Eqn.(\ref{eqn6.51}) is simply the term in square brackets on the right hand side of Eqn.(\ref{eqn6.22c}), we obtain the two equations
\begin{subequations} \label{eqn6.52}
\begin{align}
2r\left[M_{00}(\partial_{\vec{R}}h_L) + Q h_T + M_{30}\partial_{\vec{R}}h_T) \right]&=\alpha_1, \label{eqn6.52a} \\
\intertext{and}
 2r\left[M_{33}(\partial_{\vec{R}}h_T) + Q h_L + M_{03}\partial_{\vec{R}}h_L) \right]&= \alpha_2,
\label{eqn6.52b} \\
\intertext{where}
\alpha_1= \text{Tr}\big{\{} \tilde{g}_1^R( \tilde{g}_2^R \tilde{h}_2 - \tilde{h}_2\tilde{g}_2^A) +
( \tilde{g}_1^R \tilde{h}_1 - \tilde{h}_1\tilde{g}_1^A) \tilde{g}_2^A  &-  \tilde{g}_2^R( \tilde{g}_1^R \tilde{h}_1 - \tilde{h}_1\tilde{g}_1^A)- ( \tilde{g}_2^R \tilde{h}_2 - \tilde{h}_2\tilde{g}_2^A) \tilde{g}_1^A \big{\}}
 \label{eqn6.52c}, \\
\intertext{and} 
\alpha_2=  \text{Tr}\big{\{}( \tau^3 \tilde{g}_1^R( \tilde{g}_2^R \tilde{h}_2 - \tilde{h}_2\tilde{g}_2^A) +
( \tilde{g}_1^R \tilde{h}_1 - \tilde{h}_1\tilde{g}_1^A) \tilde{g}_2^A  &-  \tilde{g}_2^R( \tilde{g}_1^R \tilde{h}_1 - \tilde{h}_1\tilde{g}_1^A)- ( \tilde{g}_2^R \tilde{h}_2 - \tilde{h}_2\tilde{g}_2^A) \tilde{g}_1^A  ) \big{\}}
\label{eqn6.52d}. \\
\end{align}
\end{subequations}
We note that although the boundary conditions of Kupriyanov and Lukichev are valid for large $r$, they also give the correct boundary conditions (that of continuity of the Green's functions and distribution functions) in the limit of $r\rightarrow0$.  For arbitrary transmission of a barrier with $n$ channels, the boundary condition can be represented as\cite{nazarov}
\begin{equation}
\hat{g}_{s1} \partial_x \hat{g}_{s1}=\gamma \frac{e^2}{\pi} \underset{n}{\sum}  \;2 T_n \frac{[\hat{g}_{s1},\hat{g}_{s2}]}{4 + T_n(\hat{g}_{s1}\hat{g}_{s2} + \hat{g}_{s2}\hat{g}_{s1}- 2)},
\label{eqn6.53}
\end{equation}
where $\gamma$ is a constant factor, and $T_n$ is the transmission of the $n$th channel.  It can be seen that for a single channel with small transmission $T$, this equation reduces to the boundary condition of Kupriyanov and Lukichev.
\section{Parametrization of the quasiclassical Green's function}
The normalization Eqn.(\ref{eqn6.11a}) permits a parametrization of the quasiclassical Green's functions that is very convenient for calculations.  Eqn.(\ref{eqn6.11a}) is a matrix equation that is equivalent to the three equations (\ref{eqn5.19}) for the $s$-wave component of the Green's function.  To take into account the macroscopic phase of the superconductor, we note that  a gauge transformation that transforms the vector and scalar potentials according to Eqns.(\ref{eqn3.20}) and (\ref{eqn3.21}) transforms the field operators according to the equations
\begin{subequations} \label{eqn7.1}
\begin{align}
\hat{\psi} &\rightarrow \hat{\psi} e^{i \chi} \label{eqn7.1a} \\
\hat{\psi}^+ &\rightarrow \hat{\psi}^+ e^{-i \chi} \label{eqn7.1b} 
\end{align}
\end{subequations}
The Nambu-Gorkov Green's functions defined in Eqns.(\ref{eqn4.7}) are transformed accordingly.  For example, two components of $\tilde{G}^{\beta \alpha}_{12}$ would transform according to 
\begin{subequations} \label{eqn7.2}
\begin{align}
[\tilde{G}^{\beta \alpha}_{12}]_{11} &\rightarrow [\tilde{G}^{\beta \alpha}_{12}]_{11} e^{-i \left(\chi(\vec{r}_1) -\chi(\vec{r}_2)\right)} \label{eqn7.2a} \\
[\tilde{G}^{\beta \alpha}_{12}]_{12} &\rightarrow [\tilde{G}^{\beta \alpha}_{12}]_{12} e^{i \left(\chi(\vec{r}_1) +\chi(\vec{r}_2)\right)} \label{eqn7.2b} 
\end{align}
\end{subequations}
Making the transformation (\ref{eqn2.25a}) to mixed coordinates, and taking the limit  as $\vec{r} \rightarrow 0$, we see that the off-diagonal components of the Nambu-Gorkov Green's functions are multiplied by a phase factor $e^{i\chi(\vec{R})}$ or $e^{-i\chi(\vec{R})}$, while the diagonal components remain unchanged.  Consequently, $\Delta$ also transforms as
\begin{equation}
\Delta \rightarrow \Delta e^{i\chi(\vec{R})}
\label{7.3}
\end{equation}
Keeping this in mind, we can express $\tilde{g}_s^R$ as
\begin{equation}
\tilde{g}_s^R = 
\begin{pmatrix}
\cosh \theta & \sinh \theta e^{i \chi} \\
-\sinh \theta e^{-i \chi} & -\cosh \theta 
\end{pmatrix}
\label{eqn7.4}
\end{equation}
where $\theta$ and $\chi$ are complex functions of the energy $E$ and position $\vec{R}$.  This form satisfies the normalization condition $\hat{g}_s^R\hat{g}_s^R=\tau^0$.\footnote{One can also express $\hat{g}_s^R$ equivalently in terms of sines and cosines.}  For completeness, we also give the expression for $\tilde{g}_s^A$
\begin{equation}
\tilde{g}_s^A = 
\begin{pmatrix}
-\cosh \theta^* & -\sinh \theta^* e^{i \chi^*} \\
\sinh \theta^* e^{-i \chi^*} & \cosh \theta^* 
\end{pmatrix}
\label{eqn7.5}
\end{equation}
We now put this into the Usadel equation for $\tilde{g}^R$, Eqn.(\ref{eqn6.22a}).  Keeping in mind that the matrix for  $\tilde{\Delta}$ involves additional factors of $e^{i\chi}$ and $e^{-i\chi}$ due to the gauge transformation, the diagonal (1,1) component of this matrix equation is
\begin{subequations} \label{eqn7.6}
\begin{align}
&D \sinh^2\theta \; \partial_{\vec{R}}^2 \chi + D \sinh 2 \theta \; \partial_{\vec{R}} \chi \; \partial_{\vec{R}} \theta - 2i \Im(\Delta)\sinh\theta = 0,  \label{eqn7.6a} \\
\intertext{and the off-diagonal component (1,2) is}
&D  \partial_{\vec{R}}^2 \theta -\frac{D}{2} \sinh2\theta \;(\partial_{\vec{R}} \chi)^2 + 2Ei\sinh\theta - 2 i \Re(\Delta) \cosh\theta = 0, \label{eqn7.6b}
\end{align}
\end{subequations}
where we have used Eqn.(\ref{eqn7.6a}) to simplify Eqn.(\ref{eqn7.6b}).   Defining a current $j_s(E,\vec{R})$ by the equation
\begin{equation}
j_s(E,\vec{R})=\sinh^2\theta(E,\vec{R}) \partial_{\vec{R}}\chi(E,\vec{R}),
\label{eqn7.7}
\end{equation}
we can rewrite  Eqn.(\ref{eqn7.6a}) as
\begin{equation}
D\partial_{\vec{R}}j_s(E,\vec{R})- 2i \Im(\Delta)\sinh\theta = 0.
\label{eqn7.8}
\end{equation}
$j_s(E,\vec{R})$ is proportional to $\sinh^2\theta$, which is proportional to the square of the pair amplitude, and it is also proportional to the gauge-invariant gradient of the phase.  Consequently, it is similar in form to the conventional definition of the supercurrent, and is called the spectral supercurrent.  Indeed, Eqn.(\ref{eqn7.8}) is simply another way of writing Eqn.(\ref{eqn6.40}), and $Q$ and $j_s$ are related by
\begin{equation}
Q(E,\vec{R})=-\Im( j_s(E,\vec{R})).
\label{eqn7.9}
\end{equation}
As we noted before, $\partial_{\vec{R}}Q=0$ if $\Delta$ is purely real, and from Eqn.(\ref{eqn7.8}), it can be seen that $\partial_{\vec{R}}j_s(E,\vec{R})=0$ also if $\Delta$ is real.

Equations(\ref{eqn7.6}) form a set of coupled equations that can be solved in principle for $\theta(E, \vec{R})$ and $\chi(E, \vec{R})$.  In the case of a negligible spectral supercurrent, the equations decouple, and one needs to solve only Eqn.(\ref{eqn7.6b}).  In the limit of a bulk superconductor with a uniform real order parameter and no phase gradient, we recover the bulk value of the Green's function, Eqn.(\ref{eqn5.32}).  The differential equations must be supplemented by boundary conditions.  From Eqn.(\ref{eqn5.32}), in a superconducting reservoir, we have
\begin{equation}
\cosh \theta_{S0} = \frac{E}{\sqrt{E^2-|\Delta|^2}},
\label{eqn7.10}
\end{equation}
so that the value of $\theta$ in the superconducting reservoir is given by 
\begin{equation}
\theta_{S0}=
\begin{cases}
-i \frac{\pi}{2} + \frac{1}{2} \ln\frac{|\Delta| + E}{|\Delta| - E} & \text{if $E<|\Delta|$}, \\
\qquad \frac{1}{2} \ln\frac{E+|\Delta|}{E-|\Delta|} & \text{if $E>|\Delta|$}
\end{cases}.
\label{eqn7.11}
\end{equation}
The value of $\chi$ in a superconducting reservoir is just the macroscopic phase of the superconductor.  In a normal reservoir, $\theta=0$.  The value of $\chi$ in a normal reservoir is meaningless, of course, and any choice that results in no phase gradient is valid.

In terms of the $\theta$ parametrization, the boundary conditions of Kupriyanov and Lukichev can be expressed as
\begin{subequations}\label{eqn7.12}
\begin{align}
r \sinh \theta_1 (\partial_{\vec{R}} \chi_1) &=\sinh\theta_2 \sin(\chi_2 - \chi_1), \label{eqn7.12a} \\
\intertext{and}
r\left[\partial_{\vec{R}} \theta_1 + i \sinh \theta_1 \cosh \theta_1(\partial_{\vec{R}} \chi_1)\right]&=
 \cosh \theta_1 \sinh \theta_2 e^{i(\chi_2 - \chi_1)}-\sinh \theta_1 \cosh \theta_2 . \label{eqn7.12b}
 \end{align}
 \end{subequations}
Note that for $r=0$, the boundary conditions reduce to $\chi_1=\chi_2$ and $\theta_1=\theta_2$.  In the absence of a supercurrent, $(\partial_{\vec{R}} \chi_1)=0$, so that the equations above simplify to 
\begin{subequations}\label{eqn7.13}
\begin{align}
 \chi_1&=\chi_2, \label{eqn7.13a} \\
\intertext{and}
r(\partial_{\vec{R}} \theta_1)&=
 \sinh (\theta_2-\theta_1). \label{eqn7.13b}
\end{align}
\end{subequations}

Finally, we can write expressions for physical quantities in terms of $\theta$ and $\chi$.  These quantities can be written in terms of the $M_{ij}$
\begin{subequations} \label{eqn7.14}
\begin{align}
M_{00}&= \frac{1}{2}\left[1 + \cosh \theta \cosh \theta^* - \sinh \theta \sinh \theta^* \cosh(2 \Im (\chi))\right], \label{eqn7.14a} \\
M_{33}&= \frac{1}{2}\left[1 + \cosh \theta \cosh \theta^* + \sinh \theta \sinh \theta^* \cosh( 2\Im(\chi))\right], \label{eqn7.14b} \\
M_{03}&= \frac{1}{2}\sinh\theta \sinh \theta^* \sinh(2 \Im(\chi))), \label{eqn7.14c} \\
\intertext{and}
M_{30}&= -\frac{1}{2}\sinh\theta \sinh \theta^* \sinh(2\Im(\chi)). 
\label{eqn7.14d} 
\end{align}
\end{subequations}  
If $\chi$ is real, then the $M_{03}=M_{30}=0$, and $M_{00}$ and $M_{33}$ simplify to
\begin{subequations} \label{eqn7.15}
\begin{align}
M_{00}&=\cos^2(\Im(\theta)) \label{eqn7.15a} \\
\intertext{and}
M_{33}&=\cosh^2(\Re(\theta))  \label{eqn7.15b}
\end{align}
\end{subequations}  
These relations will be used in the next section when we discuss applying the Usadel equation to derive the transport properties of some simple device geometries.

To conclude this section, we shall write an expression for the gap in terms $\theta$.  Replacing Eqn.(\ref{eqn5.20}) for $\tilde{g}^K$ into Eqn.(\ref{eqn5.13}) for the gap, we have
\begin{equation}
\Delta=N_0 \frac{\lambda}{4} \int  dE \; [\tilde{g}_s^R \tilde{h} - \tilde{h}\tilde{g}_s^A]_{12},
\label{eqn7.16}
\end{equation}
where performing the angular average in Eqn.(\ref{eqn5.13}) gives the $s$-components of the Green's functions. Putting in $\tilde{h}$ in the form of Eqn.(\ref{eqn5.23}), we obtain
\begin{equation}
\Delta =N_0 \frac{\lambda}{4} \int  dE \; [h_L(\tilde{g}_s^R-\tilde{g}_s^A)+h_T(\tilde{g}_s^R\tau^3-\tau^3\tilde{g}_s^A)]_{12}
\label{eqn7.17}
\end{equation}
With $\tilde{g}_s^R$ and $\tilde{g}_s^A$ given by Eqns.(\ref{eqn7.4}) and (\ref{eqn7.5}), we obtain
\begin{equation}
\Delta =N_0 \frac{\lambda}{4} \int  dE \; \left[ h_L (\sinh\theta e^{i\chi} + \sinh\theta^* e^{i\chi^*})  - h_T(\sinh\theta e^{i\chi} - \sinh\theta^* e^{i\chi^*}) \right).
\label{eqn7.18}
\end{equation}
As an example, consider the case of a bulk superconductor, where $\chi=0$ and $h_T=0$.  We then have
\begin{align}
\Delta&=N_0\frac{\lambda}{2} \int dE \; h_L \Re(\sinh\theta) \nonumber \\
&=N_0\lambda \int_0^{\infty} dE   \; \tanh(E/2k_BT) \Re \left(\frac{\Delta}{\sqrt{E^2 - \Delta^2}}\right),
\label{eqn7.19}
\end{align}
which is the usual self-consistent equation for the gap.

\section{Applications of the quasiclassical equations to proximity-coupled systems}
We shall conclude our discussion of the quasiclassical theory by applying the equations that we have derived to some simple devices incorporating normal metals in close proximity with superconductors.  Since the equations of motion in the diffusive limit are in general nonlinear, solving them usually requires numerical techniques, except in the limit of large resistances between the superconductor and the normal metal, where the Usadel equation can be linearized.  We shall restrict ourselves to one-dimensional examples; these are the ones discussed most in the literature.\\

\subsection{Proximity-coupled wire}
We start with the simplest possible device, a one-dimensional normal-metal wire of length $L$ connected on one end to a superconducting reservoir and at the other end to a normal metal reservoir.  For definitiveness, let us take the superconducting reservoir to be at $x=0$ and the normal-metal reservoir at $x=L$, and let us consider first the case of a perfect SN interface, so that the interface barrier resistance parameter $r=0$.  In this geometry, there can be no supercurrent, so that $Q$ (or alternatively, $j_s$) is zero.  Furthermore, we can take the phase $\chi$ to be zero in the superconducting reservoir without loss of generality, and we note again that $\Delta=0$ in the normal metal.  Under these conditions, only $M_{00}$ and $M_{33}$ are non-zero in Eqns.(\ref{eqn6.34}), which now read
\begin{subequations} \label{eqn8.1}
\begin{align}
M_{00} (\partial_{\vec{R}} h_L)=K_1, \label{eqn8.1a} \\
\intertext{and}
M_{33} (\partial_{\vec{R}} h_T)=K_2, \label{eqn8.1b}
\end{align}
\end{subequations}
where $K_1$ and $K_2$ are constants of integration.  On integrating these equations from $x=0$ to $x=L$, we obtain
\begin{subequations} \label{eqn8.2}
\begin{align}
h_L(x=L) - h_L(x=0) = K_1 \int_0^L\frac{1}{M_{00}}dx, \label{eqn8.2a} \\
h_T(x=L) - h_T(x=0) = K_2\int_0^L\frac{1}{M_{33}} dx. \label{eqn8.2b} 
\end{align}
\end{subequations}
To calculate the conductance of the normal-metal wire in the linear approximation, we apply a small voltage $V$ on the normal-metal reservoir, keeping the superconducting reservoir at $V=0$.  If we consider the second equation, $h_T(x=0)=0$, and expanding $h_T(x=L)$ in a Taylor's expansion to first order, we obtain from Eqn.(\ref{eqn8.2b})
\begin{equation}
K_2=\frac{eV}{2k_BT\cosh^2\left(\frac{E}{2k_BT}\right)}
\left[\int_0^L\frac{1}{M_{33}(E,x)} dx\right]^{-1}.
\label{eqn8.3}
\end{equation}
The electric current in the linear response regime can then be obtained from Eqn.(\ref{eqn6.35})
\begin{equation}
j=\frac{N_0 e^2 V D}{2k_B T} \int dE \frac{1}{ \cosh^2(E/2k_BT)} \left[\int_0^L \frac{1}{M_{33}(E,x)} dx\right]^{-1}
\label{eqn8.4}
\end{equation}
There are two differences between this equation and the equivalent equation for a normal metal in the classical regime (Eqn.(\ref{eqn1.10a}) that we derived earlier.  First, the equation above involves an integral over energy and position.  One can define a energy and position dependent electrical diffusion coefficient
\begin{equation}
D_3(E,x)=DM_{33}(E,x)
\label{eqn8.5}
\end{equation}
instead of the constant diffusion coefficient $D$ for Eqn(\ref{eqn1.10a}).  Second, the current in Eqn.(\ref{eqn8.4}) does not involve temperature differentials.  Indeed, if one assumes that the superconducting reservoir is at a temperature $T$, and the normal reservoir at a temperature $T+\Delta T$, and expand $h_T(x=L)$  to first order in $\Delta T$, the terms involving $\Delta T$ cancel, so that there is no term proportional to $\Delta T$.  As a consequence, thermoelectric phenomena cannot be described using the quasiclassical equations, and an extension to the theory is required to take them into account.\footnote{See, for example, F. Wilhelm, PhD Thesis, Universit\"at Karlsruhe, 2000}

From Eqn.(\ref{eqn8.4}), one can define a spectral or energy dependent conductance of the wire 
\begin{equation}
G(E)=G_N \left[\int_0^L \frac{1}{M_{33}(E,x)} dx\right]^{-1},
\label{eqn8.6}
\end{equation}
where $G_N$ is the normal state conductance of the wire.  The total conductance is then
\begin{equation}
G=\int dE \frac{G(E)}{2k_BT\cosh^2(E/2k_BT)}
\label{eqn8.7}
\end{equation}

\begin{figure}[ht]
\center{\includegraphics[width=8cm]{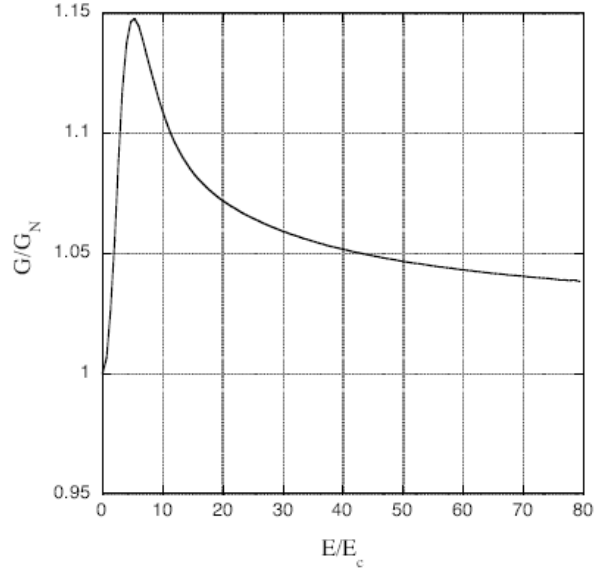}}
\caption{Spectral conductance $G(E)$ of a one-dimensional wire of length $L$ as a function of energy $E$, normalized to the correlation energy $E_c$.  The barrier interface parameter $r$=0, and the gap is set to be $\Delta=1000 E_c$.}
\label{fig4}
\end{figure}

Figure 4 shows the results of a numerical calculation of $G(E)/G_N$ as a function of the normalized energy $E/E_c$.  The normalization factor $E_c= D/L^2$ is called the correlation energy or Thouless energy (from the theory of disordered metals, where it also occurs),\cite{altshuler} and is dependent on the length $L$ of the wire. At high energies, $G(E)$ approaches its normal state value, as expected.  As the energy is lowered, the conductance increases, as might be expected in a proximity-coupled normal metal.  However, instead of continually increasing as the energy is reduced, it reaches a maximum of around $1.15 \; G_N$ at an energy of $E\simeq5 E_c$, and then decreases, reaching its normal state value at $E=0$.  This nonmonotonic behavior of the conductance is called reentrance, and has been observed in experiments by a number of groups.\cite{charlat}  It should be emphasized that the relevant energy scale where the maximum in conductance is observed is set not by the gap $\Delta$ of the superconductor, but by $E_c$, which itself depends inversely on the square of the length of the sample $L$.  Hence in very long or macroscopic samples, the energy (and correspondingly, the temperature) at which the minimum would occur is far below the experimentally accessible range, and one regains the monotonic behavior expected from the simple Ginzburg-Landau theory of de Gennes.\cite{degennes}

The temperature dependent conductance $G(T)$ can be obtained from $G(E)$ using Eqn.(\ref{eqn8.7}); the result of this calculation, plotted in terms of the normalized resistance $R(T)/R_n$, is shown in Fig. 5.  In obtaining this plot, we have used a value of $\Delta=32 E_c$, corresponding to a weak-coupling transition temperature of $T_c=1.764 \Delta/k_B$, typical parameters for Al films.  We have also assumed that the gap is temperature dependent.  Like $G(E)$, $R(T)$ is also non-monotonic, with a minimum at some intermediate temperature.  We would expect the minimum in $R(T)$ to be around $T\simeq5 E_c/k_B$, based on the behavior of $G(E)$.  However, the temperature dependence of the superconducting gap modifies this behavior, so that the minimum in resistance occurs at a somewhat higher temperature when the interface between the normal metal and the superconductor is perfect.  Figure 5 shows additional curves corresponding to progressively increasing values of the interface barrier parameter $r$.      
Increasing the resistance of the NS interface decreases the leakage of superconducting correlations from the superconductor into the normal metal, and consequently results in a smaller increase in the conductance of the proximity-coupled normal metal.  In addition, the temperature $T_{min}$ at which the minimum in resistance occurs is also shifted down.   

\begin{figure}[ht]
\center{\includegraphics[width=8cm]{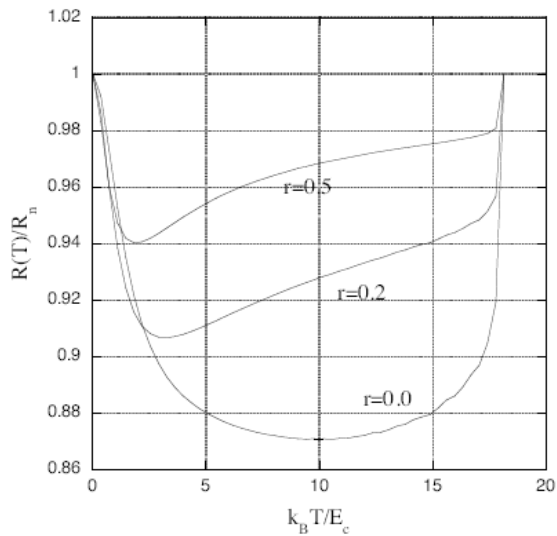}}
\caption{Temperature dependent resistance $R(T)$ normalized to the normal state resistance $R_N$, as a function of the temperature $T$ normalized to $E_c$, for different values of the interface resistance parameter $r$.  The gap is set to be $\Delta=32 E_c$.}
\label{fig5}
\end{figure}

From Eqn.(\ref{eqn6.38}), the normalized density of states $N(E)$ can be expressed in terms of $\theta$ as
\begin{equation}
N(E)=\cosh\left(\Re(\theta)\right)\cos\left(\Im(\theta)\right).
\label{eqn8.8}
\end{equation}
Figure 6 shows the density of states as a function of energy and position along the wire of length $L$.  There is a proximity-induced decrease of $N(E)$ near the superconducting reservoir.  In fact, at the NS interface, there is a divergence in $N(E)$ at the gap energy, and it goes to zero at $E=0$, just as one would expect for a superconductor.  However, unlike a superconductor, it is not strictly zero for $E<\Delta$, but still has a finite amplitude.  As one moves away from the NS interface into the proximity-coupled normal wire, both the amplitude of this effective gap and the divergence are smoothly reduced, so that at the normal reservoir, one recovers the normal density of states.  This position dependent variation of the density of states has been observed in experiments.\cite{gueron}  

\begin{figure}[ht]
\center{\includegraphics[width=8cm]{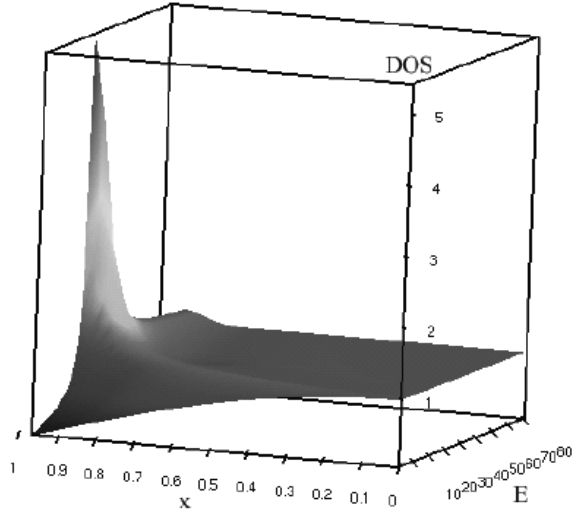}}
\caption{Density of state $N(E,X)$ normalized to the normal state density of states  $N_0$ for a one dimensional wire of length $L$, as a function of $E/E_c$ and position $x$ along the wire.  The superconducting reservoir is at $x=1.0$, and the normal reservoir is at $x=0$.  The density of states is suppressed at low energies near the superconducting reservoir.}
\label{fig6}
\end{figure}

In our analysis above of the proximity effect in a normal metal coupled to a superconductor, we have ignored the effects of electron decoherence on the proximity correction.  Phase coherence is essential to observing the proximity effect; if the phase coherence length $L_{\phi}$ is less than the length of the sample $L$, a finite spatial cutoff of the proximity effect is introduced.  Phenomenologically, this can be taken into account by saying that the length of the wire is now $L_{\phi}$ instead of $L$.  Since $L_\phi$ is typically of the order of a few microns even at low temperatures, this sets the dimensions of the samples that are required to observe this mesoscopic proximity effect.

While $L_\phi$ sets the upper cutoff for observing the proximity effect, a second relevant length scale for the problem can be obtained by considering the length at which $E_c$ is equal to $k_B T$.  This length is
\begin{equation}
L_T = \sqrt{\frac{\hbar D}{k_B T}},
\label{eqn8.9}
\end{equation}
where $L_T$ is called variously the thermal diffusion length or the Thouless length, again from the theory of disordered metals, where it also occurs.\cite{altshuler}  (We put in here explicitly the factor of $\hbar$.)  In fact, here it is simply the diffusive form of the superconducting coherence length in the normal metal, familar from the de Gennes/Ginzburg-Landau theory of the proximity effect,\cite{degennes} which in the clean limit is given by
\begin{equation}
\xi_N = \frac{\hbar v_F}{ k_B T}.
\label{eqn8.10}
\end{equation}
At low temperatures, when $L_T$ is longer than $L$, the superconducting correlations induced in the normal metal extend throughout its length.  At higher temperatures, they are restricted to a region of length $L_T$ near the superconductor.  $L_T$ is also on the order of a few microns in typical metallic samples in accessible temperature regimes, so it also sets a limit on the dimensions of the samples in which one can see a proximity effect. 

To calculate the thermal conductance of the wire, we proceed from Eqn.(\ref{eqn8.2a}).  We now consider a small temperature differential $\Delta T$ applied across the wire.  Expanding $h_L$ in a first-order Taylor's expansion, as we did for $h_T$, we obtain
\begin{equation}
K_1=-\frac{E\Delta T}{2k_BT^2\cosh^2\left(\frac{E}{2k_BT}\right)}\left[\int_0^L\frac{1}{M_{00}(E,x)} dx.\right]^{-1}
\label{eqn8.11}
\end{equation}
We then obtain from Eqn.(\ref{eqn6.36})
\begin{equation}
j_{th}=-\frac{N_0  D \Delta T}{2k_B T^2} \int dE \frac{E^2}{ \cosh^2(E/2k_BT)} \left[\int_0^L \frac{1}{M_{00}(E,x)} dx\right]^{-1}
\label{eqn8.12}
\end{equation}
As with the electrical conductance, we can define a thermal diffusion coefficient
\begin{equation}
D_0(E,x)=DM_{00}(E,x),
\label{eqn8.13}
\end{equation}
and a spectral thermal conductance
\begin{equation}
G_{th}(E)=G_{thN} \left[\int_0^L \frac{1}{M_{00}(E,x)} dx\right]^{-1},
\label{eqn8.14}
\end{equation}
where $G_{thN}$  is related to the normal state electrical conductance by  Eqn.(\ref{eqn1.17})
\begin{equation}
G_{thN}=G_N \frac{\pi^2}{3} \frac{k_B^2}{e^2}T.
\label{eqn8.15}
\end{equation} 
Finally, the thermal conductance itself is given by an integral over energy
\begin{equation}
G_{th}=
\frac{3}{\pi^2}\frac{1}{2(k_BT)^3}\int dE \frac{E^2G_{th}(E)}{\cosh^2(E/2k_BT)}.
\label{eqn8.16}
\end{equation}
Of course, as noted by Andreev,\cite{andreev} the thermal conductance of a normal metal wire sandwiched between a normal-metal reservoir on one end and a superconducting reservoir on the other end must vanish, since the superconductor acts as a thermal insulator, so that no thermal current can flow through the device as a whole.  However, one may consider a normal-metal wire with the superconducting reservoir connected off to one side, so that it does not block the flow of thermal current through the proximity-coupled wire.  One may then consider the thermal conductance of the normal metal wire itself.  Figure 7 shows the thermal conductance of this geometry, as a function of temperature, for different transmissivities of the interface barrier.  The thermal conductance shows a monotonic decrease as $T$ is lowered below $T_c$, although there are no distinct features at any particular temperature, unlike for the electrical conductance.  In a superconductor, the exponential decrease in the thermal conductivity is associated with the opening of the gap in the quasiparticle density of states, since it is the quasiparticles that carry the thermal current.  Noting the decrease in the density of states in the proximity-coupled normal metal wire, shown in Fig. 6, it is not surprising that this system will also show a decrease in the thermal conductance.  The thermal conductance of the wire is strongly dependent on the transmission of the NS interface, characterized by the parameter $r$, and approaches the normal state thermal conductance as $r$ increases.

\begin{figure}[ht]
\center{\includegraphics[width=8cm]{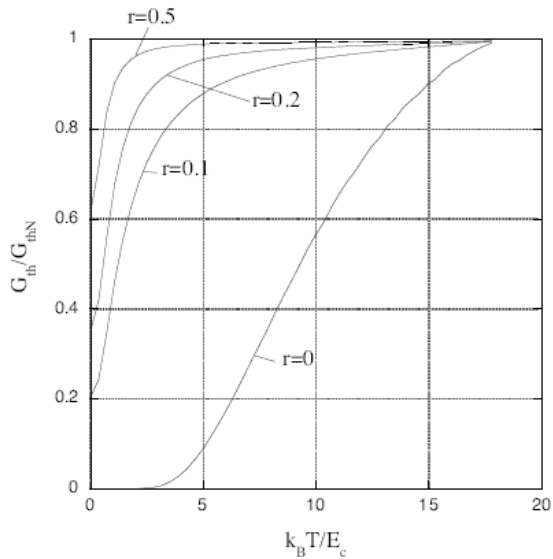}}
\caption{Thermal conductance $G_{th}$ of a normal wire connected to a superconducting reservoir on one end, and a normal metal reservoir on the other, as a function of the normalized temperature $T$, for a number of different values of the interface barrier parameter $r$.  The gap is set to be $\Delta=32 E_c$, corresponding to a transition temperature of $T=18.14 E_c$.}
\label{fig7}
\end{figure}

We note here again that, in our current approximation, a small voltage drop across the S/n-wire/N device will not result in a contribute to a thermal current through the system, since any terms proportional to voltage in the expansion of $h_L$ will cancel.  This is the converse of the case for the electrical current, where a small temperature drop did not contribute to the electrical current, emphasizing again that the conventional quasiclassical approximation cannot take into account thermoelectric effects.

\subsection{Superconductor-metal bilayer}
Instead of a normal reservoir on one side of the wire, if we consider a wire of length $L$ connected only on one end to a superconducting reservoir (with the other end open), then a true gap in the density of states opens up in the proximity-coupled  normal metal.  The magnitude of the gap is related to $E_c$; hence, one can consider the proximity-coupled normal-metal in this case as a superconductor with a gap of $E_c$.

If the superconductor is not a reservoir, but a thin layer itself, then one will suppress superconductivity in the superconducting layer due to the proximity of the normal metal, an inverse proximity effect.  The suppression of superconductivity is expected to reduce the gap in the superconductor.  It is an interesting exercise to calculate the transition temperature of the bilayer in the quasiclassical approximation.  We shall loosely follow here the treatment given by Martinis \textit{et al.}\cite{martinis} and Gu\'eron.\cite{gueron2} 

Let the thickness of the superconductor be $t_S$, and the thickness of the normal metal $t_N$.  We take the origin, $x=0$, at the NS interface, and the superconductor extends from $x=-t_S$ to $x=0$, and the normal metal from $x=0$ to $x=t_N$.  Near the transition, the order parameter in the superconductor is small, so we may the small $\theta$ limit of Eqn.(\ref{eqn7.6b}).  Since the phase is not important in this problem, we take $\chi=0$.  The resulting equation is 
\begin{equation}
D \partial_x^2 \theta  + 2 E i \theta - 2 i \Delta = 0,
\label{eqn8.17}
\end{equation}
where we have assumed that the gauge is chosen so that $\Delta$ is real.  Let us assume that $\theta$ at $x=0$ in the superconductor is $\theta_{0S}$, and that variations of $\theta$ about this mean value are small.  Under these conditions, we can also assume that the gap $\Delta$ in the superconductor is uniform.  We can then expand $\theta_S$ to second order in $x$
\begin{equation}
\theta_S = \theta_{0S} + a x + b x^2.
\label{eqn8.18}
\end{equation} 
Now, at $x=-t_S$, (and also at $x=t_N$), we have a vacuum interface, where $\partial_x\theta=0$.  Hence
$a=2bt_S$, and from the differential equation (\ref{eqn8.17}) taken at $x=0$, $b=(i/D)(\Delta-E\theta_{0S})$, so that 
\begin{equation}
\theta_S = \theta_{0S} + \frac{i}{D}(\Delta-E\theta_{0S})(2t_Sx + x^2).
\label{eqn8.19}
\end{equation} 
Similarly
\begin{equation}
\theta_N = \theta_{0N} + \frac{i}{D}E\theta_{0N}(2t_Nx - x^2).
\label{eqn8.20}
\end{equation} 
From the boundary condition Eqn.(\ref{eqn7.13b}), we have the two equations
\begin{subequations} \label{eqn8.21}
\begin{align}
\frac{2irt_S}{D}(\Delta-E\theta_{0S})&=\theta_{0N}-\theta_{0S} \label{eqn8.21a}, \\
\frac{2irt_N}{D}E\theta_{0N}&=\theta_{0N}-\theta_{0S} \label{eqn8.21b}. 
\end{align}
\end{subequations}
Solving this pair of equations for $\theta_{0S}$, we  have
\begin{equation}
\theta_{0S}=\frac{\Delta}{E}\frac{D^2t_S(t_S+t_N) + 4 r^2 t_N^2 t_S^2 -2ir Dt_n^2}{D^2(t_s+t_N)^2 + 4 r^2 t_S^2 t_N^2}.
\label{eqn8.22}
\end{equation}
Putting this into Eqn.(\ref{eqn7.19}) for the gap, with $T=T_c$, and with the approximation that $\sinh(\theta)\simeq\theta$, we obtain
\begin{equation}
1=N_0 \frac{\lambda}{2} \int \frac{dE}{E} \tanh(E/2k_B T_c) \left[1 - \frac{t_N(t_S+t_N)}{(t_S+t_N)^2 + (4r^2/D^2)t_S^2 t_N^2} \right].
\label{eqn8.23}
\end{equation}
The first term in the square brackets gives the bare transition temperature $T_{c0}$ of the superconducting film, and the second term corresponds to the corrections associated with the inverse proximity effect.  For a perfect interface, with $r=0$, the suppression of $T_c$ is directly proportional to the normal fraction of the bilayer, and $T_c\rightarrow0$ as $t_N$ increases.  For a highly resisitive barrier ($r\rightarrow \infty$), the second term in the square brackets goes to zero, so that there is little effect of the normal metal film on $T_c$ of the superconducting film, as expected.  

\subsection{The SNS junction and Andreev interferometers}
As our final example of the application of the quasiclassical equations of superconductivity, we consider the case of a dirty SNS junction.  The model we consider is a normal metal wire of length $L$ sandwiched between two superconducting reservoirs.  As is well known, the application of a phase difference between the two superconducting reservoirs will result in the flow of a supercurrent through the normal wire.  The phase difference can be applied, for example, by connecting one of the superconducting reservoirs to the other, thereby forming a loop with two different arms, one superconducting and one normal.  This configuration is commonly called an Andreev interferometer.  The phase between the two superconducting reservoirs can be varied by applying an Aharonov-Bohm type magnetic flux through the area of the loop; in this respect, we put together all contributions in the gauge-invariant phase $\chi$.  Due to the single-valued nature of the wave functions, a phase factor of $2 \pi \Phi /\Phi_0$ is picked up in going aroung the loop, where $\Phi$ is the magnetic flux threading the Andreev interferometer, and $\Phi_0 = h/2e$ is the superconducting flux quantum.  In a superconductor, the supercurrent $I_S$ that is generated is directly proportional to the phase gradient; if $I_S$ is small compared to the critical current $I_c$, the phase dropped across the superconductor will be small.  Since $I_c$ of the superconducting part of the Andreev interferometer is so much greater than the critical current of the proximity coupled normal-metal wire, most of the phase change will occur across the length $L$ of the normal metal wire.

This fact allows us to map the Andreev interferometer that is coupled with an Aharonov-Bohm flux $\Phi$ to a SNS system with a phase difference $\phi = 2 \pi \Phi /\Phi_0$ across it.  In terms of our model, we consider the superconductors to be reservoir; this means applying a boundary condition for the gauge-invariant phase $\chi$ at the superconducting reservoirs.  For our purposes, we apply this boundary condition anti-symmetrically, with a phase $\chi_L=-\pi \Phi/\Phi_0$ at the superconducting reservoir at $x=0$, and $\chi_R=\pi \Phi/\Phi_0$ at the superconducting reservoir at $x=L$.  We then must solve Eqns.(\ref{eqn7.6}) in the normal-metal wire for $\theta$ and $\chi$, with $\Delta=0$, subject to the boundary condition for $\chi$ noted above, and the boundary condition $\theta=\theta_{S0}$ (where $\theta_{S0}$ is given by Eqn.(\ref{eqn7.11})) at both $x=0$ and $x=L$.  In general, both $\theta$ and $\chi$ are complex functions of $x$ and $E$, and the solution of Eqns.(\ref{eqn7.6}) must be done numerically.  

\begin{figure}[ht]
\center{\includegraphics[width=12cm]{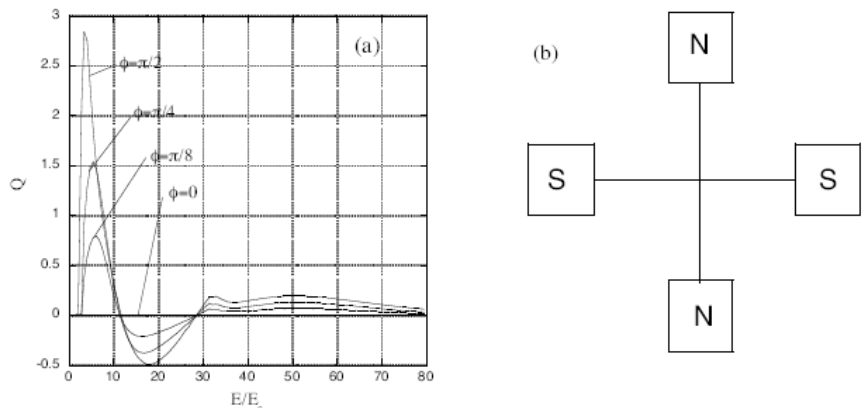}}
\caption{(a) Spectral supercurrent $Q$ in a normal wire between two superconducting reservoirs, as a function of the normalized energy $E/E_c$, for different values of the phase difference $\phi$ between the superconducting reservoirs.  (b)  Geometry of an Andreev interferometer, essentially a cross with one arm connected to superconducting reservoirs, and the other arm connected to normal reservoirs.}
\label{fig8}
\end{figure}

Following Yip,\cite{yip} we consider first the supercurrent $Q$, given by Eqn.(\ref{eqn7.9}).  Some insight into the contributions to $j_s$ can be gained by looking again at the case of a bulk superconductor.  From Eqn.(\ref{eqn6.46}), the major contribution to $j_s$ comes from energies near the gap.  For a long proximity wire with $E_c<\Delta$, however, the major contribution comes from energies of order $E_c$.  Figure 8(a) shows a plot of $Q$ as a function of energy for various values of the phase difference $\phi$ between the two superconducting reservoirs, for the case of a perfectly transparent interface, and $\Delta=32 E_c$.  Of course, for zero phase difference, the supercurrent vanishes.  As $\phi$ is increased from zero, there is a peak in $Q(E)$ at $E\simeq6E_c$.  This peak moves down in energy as $\phi$ increases.  At larger values of $E$, $Q$ becomes negative.  For shorter wires, this region of negative $Q$ is less prominent.  The total supercurrent is given by the second term in Eqn.(\ref{eqn6.35})
\begin{equation}
J_s=eN_0D \int dE \; Q(E) h_L(E),
\label{eqn8.24}
\end{equation}
and therefore depends also on the distribution of quasiparticles.  Any change in this distribution will affect the supercurrent.  For example, the distribution can be changed by increasing the temperature, which has the result of decreasing the supercurrent.  As we demonstrated in the introduction, a non-equilibrium quasiparticle distribution can also be generated by injecting a normal current into the proximity wire in the SNS geometry by attaching two additional leads to the center of the normal wire, forming a normal cross, with two of the wires attached to superconducting reservoirs, and the two other wires attached to normal reservoirs, as shown in Fig. 8(b).  The current in the SNS junction is then a function of the current injected between the normal reservoirs, and the supercurrent can even change sign depending on magnitude of the injected normal current.\cite{yip,wilhelm2}  This effect has been observed in recent experiments.\cite{baselmans}

Due to long-range phase coherence, the Green's function in the arms of the cross attached to the normal reservoirs will also depend on the phase difference between the two superconducting reservoirs in the structure shown in Fig. 8(b).  Consequently, the electrical conductance measured between the two normal reservoirs will also be a periodic function of the phase difference between the two superconducting reservoirs.  Experimentally, the conductance of such Andreev interferometers have been found to oscillate periodically with an applied external flux, with a fundamental period of $\phi_0=h/2e$.\cite{pothier,petrashov,denhartog}  Similar oscillations are also expected in the thermal conductance as well.  Periodic oscillations are also observed in the thermopower of Andreev interferometers,\cite{eom} although these thermopower oscillations cannot be described within the framework of the current quasiclassical theory.

\section{Summary}
The quasiclassical theory of superconductivity has proved to be a powerful tool for the quantitative description of long-range phase coherent phenomena in diffusive proximity coupled systems.  As we have shown, the linear electrical and thermal conductance of complicated devices incorporating normal and superconducting elements can be calculated in principle, although the solutions frequently involve numerical techniques.  Extension to the nonlinear regime, with finite voltages across the normal reservoirs, is also conceptually straightforward, although numerically challenging.  

Application of finite voltages to the superconducting elements is trickier, as it involves time dependent evolution of the phase, and is only beginning to be examined theoretically.  Finally, the quasiclassical theory for diffusive systems, in its present form, does not deal at all with thermoelectric phenomena.  Extensions to incorporate thermoelectric effects in the theoretical framework have been attempted,\cite{wilhelm3} but still require further work to be complete. 


\end{document}